\documentclass[reprint,amsmath,amssymb,aps,superscriptaddress,prc]{revtex4-1}

\usepackage{amsmath}
\usepackage{graphicx}
\usepackage{bm}
\graphicspath{{Figures/}}
\usepackage[caption=false,subrefformat=parens]{subfig}
\usepackage{hyperref}

\newcommand{\beq}{\begin{equation}}
\newcommand{\enq}{\end{equation}}
\newcommand{\beqn}{\begin{eqnarray}}
\newcommand{\enqn}{\end{eqnarray}}
\newcommand{\al}{\alpha}
\newcommand{\be}{\beta}
\newcommand{\ga}{\gamma}
\newcommand{\de}{\delta}
\newcommand{\ep}{\epsilon}

\newcommand{\om}{\omega}
\newcommand{\vep}{\varepsilon}
\newcommand{\f}[2]{\frac{#1}{#2}}
\newcommand{\h}[1]{\hat{#1}}
\newcommand{\dg}{\dagger}
\newcommand{\nn}{\nonumber}
\renewcommand{\d}{\textrm d}

\begin{document}

\title{Correlated density-dependent chiral forces for infinite-matter calculations \\ within the Green's function approach}

\author{Arianna Carbone}
\altaffiliation[Present address: ]{Institut f\"ur Kernphysik, Technische Universit\"at Darmstadt, 64289 Darmstadt, Germany and ExtreMe Matter Institute EMMI, GSI Helmholtzzentrum f\"ur Schwerionenforschung GmbH, 64291 Darmstadt, Germany}
\email{email address: arianna@theorie.ikp.physik.tu-darmstadt.de}
\affiliation{Departament d'Estructura i Constituents de la Mat\`eria and Institut de 
Ci\`{e}ncies del Cosmos, Universitat de Barcelona, E-08028 Barcelona, Spain}

\author{Arnau Rios}
\email{Email address: a.rios@surrey.ac.uk}
\affiliation{Department of Physics, Faculty of Engineering and Physical Sciences, University of Surrey, Guildford, Surrey GU2 7XH, United Kingdom}

\author{Artur Polls}
\email{Email address: artur@ecm.ub.edu}
\affiliation{Departament d'Estructura i Constituents de la Mat\`eria and Institut de 
Ci\`{e}ncies del Cosmos, Universitat de Barcelona, E-08028 Barcelona, Spain}

\date{\today}

\begin{abstract}

The properties of symmetric nuclear and pure neutron matter are investigated within an extended self-consistent Green's function method that includes the effects of three-body forces. We use the ladder approximation for the study of infinite nuclear matter and incorporate the three-body interaction by means of a density-dependent two-body force. This force is obtained via a correlated average over the third particle, with an in-medium propagator consistent with the many-body calculation we perform. We analyze different prescriptions in the construction of the average, and conclude that correlations provide small modifications at the level of the density-dependent force. Microscopic as well as bulk properties are studied, focusing on the changes introduced by the density dependent two-body force. The total energy of the system is obtained by means of a modified  Galitskii-Migdal-Koltun sum rule. Our results validate previously used uncorrelated averages and extend the availability of chirally motivated forces to a larger density regime. 
\end{abstract}


\maketitle

\section{Introduction}
\label{intro}
State-of-the-art \emph{ab initio} nuclear many-body theories necessarily include both two-body (2B) and three-body (3B) interactions. The importance of three-nucleon interactions was first recognized with the pioneering work of Fujita and Miyazawa, which identified two-pion exchange (TPE) as an essential underlying process \cite{Fujita1957}. The inclusion of three-nucleon forces (3NFs) becomes mandatory to avoid the underbinding of light nuclei and to elude the saturation of nuclear matter at too-high densities \cite{Carlson1983}. Several models of 3NFs have been devised in the past few decades \cite{McKellar1968,*Brown1969,*Lagaris1981,*Grange1989,*Pieper2001}. In some of these models, the TPE part of the 3NF has been complemented with additional phenomenological terms, for instance, the repulsive terms appearing in the 3NF Urbana interactions \cite{Pieper2001}. While their inclusion is empirically reasonable, it can be hard to justify from a theoretical point of view. Consequently, systematic errors become difficult to quantify and the predictive power of the theory is adversely affected. Furthermore, \emph{ab initio} methods should be constructed considering a unified description of 2B and 3B interactions, avoiding if at all possible \emph{ad hoc} terms. In this context, we exploit the consistent framework provided by chiral nuclear forces to study many-body systems with a nonperturbative self-consistent Green's function (SCGF) approach \cite{Dickhoff2004}.

Chiral effective field theory (EFT) provides a consistent picture of nuclear interactions; see Refs.~\cite{Epelbaum2009,Machleidt2011} for recent reviews. The approach is based on chiral perturbation theory, i.e., a low-energy expansion of quantum chromodynamics (QCD) \cite{Epelbaum2012}. The expansion is based on a power counting, which generally involves a low-momentum scale over a high-energy scale, usually taken as the chiral symmetry breaking scale. The expansion is constructed as to be consistent with the symmetries of the underlying quantum theory, QCD. The high-energy physics is incorporated in low-energy constants (LECs), which need to be fit to low-energy hadronic and nuclear structure properties  \cite{Epelbaum2009,Machleidt2011}. A tremendous advantage of the resulting expansion is the fact that two- and many-body nuclear forces are organized according to the same consistent power counting. 3NFs, for instance, appear at next-to-next-to-leading order (N2LO) in the chiral expansion, whereas four-nucleon forces only appear at fourth order (N3LO). In this sense, a full calculation using chiral forces should be performed considering all the $n$-body forces which are present at a given order in the chiral expansion. The inclusion of many-body forces, other than the 2B ones, is unavoidable when dealing with chiral nuclear interactions \cite{Bogner2005,Bogner2010}. 

Most nuclear many-body formalisms were originally employed with two-nucleon forces (2NFs) alone. The majority of these methods have been extended, in one way or another, to include 3N interactions. For finite systems, the SCGF approach \cite{Cipollone2013}, the Gorkov-Green's function method \cite{Soma2014}, the no-core shell model \cite{Lisetskiy2008,Roth2012}, the in-medium similarity renormalization group \cite{Hergert2013} or the coupled-cluster formalisms \cite{Hagen2007,Binder2013}, have recently been developed to treat 3B forces. One of the motivations underlying all these developments is the modified shell-model results of Ref.~\cite{Otsuka2010}, which demonstrated the importance of 3NFs in the reproduction of the drip line in oxygen isotopes. Recently, nuclear EFT calculations on the lattice have provided solid benchmark results for \emph{ab initio} calculations in the light to medium mass range \cite{Laehde2014}.

For infinite matter, similar extensions have been devised over the years. In bulk matter the 3B forces have been mostly included in a density-dependent 2B form \cite{Zuo2002,Bogner2005,Li2006,Hebeler2010Jul,Lovato2011,Lovato2012,Yin2013}. In particular, calculations using the SCGF method for both symmetric nuclear matter (SNM) and pure neutron matter (PNM), have already been presented in Refs.~\cite{Soma2008,*Soma2009}. The practical advantages of using a density-dependent 2B force constructed from 3NFs are numerous. First, the matrix elements can essentially be precomputed at the start of the calculation if they have been built from uncorrelated averages. Second, and most important, the use of the density dependent force avoids the problem of solving the corresponding Faddeev-type equations. In principle the density-dependent force can be added directly to the original 2NF before performing many-body calculations. However, in diagrammatic calculations, this is not a correct prescription, as we have discussed previously \cite{Carbone2013Oct,Carbone2014}. 

Chiral forces have been implemented at the 2B and 3B level in perturbative as well as in nonperturbative infinite matter calculations. Perturbative calculations have been performed in both SNM \cite{Hebeler2011} and PNM \cite{Hebeler2010Jul}, up to third-order in the ladder diagrams for the former and second order for the latter, using an evolved N3LO 2NF, via renormalization group techniques \cite{Bogner2010}, plus a density-dependent N2LO 2NF. Calculations up to N3LO  in the chiral expansion with 2B, 3B and four-body forces have been presented for PNM in Ref.~\cite{Krueger2013}. We note, however, that pure N3LO many-body terms have only been included at the Hartree-Fock level. Up-to-third-order calculations in the energy expansion were presented for PNM in Ref.~\cite{Coraggio2013}, and later for SNM including particle-hole diagrams \cite{Coraggio2014}. Furthermore, second order calculations in many-body perturbation theory have been recently presented at finite temperature \cite{Wellenhofer2014}. 

Nonperturbative calculations using 2B and 3B chiral forces have been carried out with several many-body approaches. Initial Brueckner-Hartree-Fock calculations with chiral interactions found anomalously attractive results \cite{Li2012}. In contrast, satisfactory results in the Brueckner-Hartree-Fock approach have been presented in Ref.~\cite{Kohno2013}. These agree qualitatively with our SCGF analysis, presented in Ref.~\cite{Carbone2013Oct}. The Fermi-hyper-netted-chain method as well as the auxiliary field diffusion Monte Carlo method have been complemented in the 3B sector using a new generation of chirally inspired 3NFs in coordinate space \cite{Lovato2012}. Coupled-cluster calculations in infinite nucleonic matter have also been performed, exploiting a newly optimized version of the 2NF at third order in the chiral expansion (N2LOopt) \cite{Baardsen2013,Hagen2014}. Promising quantum Monte Carlo calculations have been obtained with chiral local 2NF at N2LO  \cite{Gezerlis2013,*Gezerlis2014} and with chiral non-local N2LOopt forces \cite{Roggero2014}. Furthermore, quantum Monte Carlo simulations of neutrons interacting on the lattice have been recently performed including N3LO 2NFs together with 3NFs at N2LO \cite{Wlazlowski2014}.

We have recently extended the SCGF formalism \cite{Dickhoff2004} to incorporate 3B forces from the outset \cite{Carbone2013Nov}. In short, the extension involves a new diagrammatic expansion, for which several structures can be subsumed using effective interactions. Nonperturbative resummation schemes can be developed, and those employed with 2B interactions can, in general, be extended to account for 3NFs. The many-body method is self-consistent, in that an iterative solution of the Dyson's equation is needed to find a solution for the in-medium propagator. In other words, the Green's function (GF) which describes a propagating particle through the medium is determined by an interaction which takes into account the medium itself. Conversely, for the construction of the in-medium interaction, the knowledge of the propagator is required. 

The successful extension of the SCGF formalism to include 3B forces has been implemented in a variety of systems. Recent calculations have found good agreement between theoretical and experimental ground state properties of nitrogen, oxygen and fluorine isotopes  \cite{Cipollone2013}. Moreover, the extension to 3B forces has been implemented in the Gorkov-Green's function method \cite{Soma2013Nov} to access mid-mass open-shell nuclei \cite{Soma2014}. Results in finite nuclei have been obtained using a local version of the N2LO 3B force \cite{Navratil2007Nov}. In infinite matter, we have presented calculations for SNM \cite{Carbone2013Oct}, analyzing the modifications introduced by 3B forces on both microscopic and macroscopic properties. In our study, we have confirmed the important role played by 3NFs in the saturation mechanism of nuclear matter. In fact, 3NF-induced repulsion provides more realistic values for the saturation energy and density. Throughout our calculations a non-local 3NF at N2LO is implemented \cite{vanKolck1994,Epelbaum2002Dec2}.

In our previous publication, however, 3B interactions were incorporated via a density-dependent two-body force based on an uncorrelated average over the third particle. In the following, we explore a natural improvement of this average, calculating a correlated average consistent with the many-body calculation. We provide comparisons with our previous results and with other groups' calculations. Furthermore, and for the first time within the extended SCGF formalism, results for PNM are presented. 

We will exploit a density-dependent 2NF constructed from a correlated contraction of diagrams appearing at N2LO in the 3B sector \cite{vanKolck1994,Epelbaum2002Dec2}. We point out that this average differs with respect to previous approaches \cite{JWHolt2010,Hebeler2010Jul} in that it takes into account the correlations characterizing the system. In other words, the correlated momentum distribution, obtained at each stage of the iterative many-body calculation, is consistently used to calculate the averaged 3B potential. This density-dependent force will be complemented with the 2NF N3LO Entem-Machleidt potential \cite{Entem2003}. We are well aware that, because we are using different orders in the chiral expansion for the 2B and 3B sector, our calculations  are inconsistent in the way they treat chiral nuclear forces. In consequence, we will also present an exploratory study implementing the N2LOopt interaction at the 2B level \cite{Ekstroem2013}.
	
This paper is divided as follows. Formal aspects are discussed in Sec.~\ref{sec2a}, where a brief revision of the extended ladder approximation to perform calculations in infinite matter is presented. Section~\ref{sec2b} will be mainly dedicated to the analysis of the correlated density-dependent force. Furthermore, we will explore different manners of performing the third-particle average. Section~\ref{sec3} will focus on the study of SNM, to complement the work already presented in Ref.~\cite{Carbone2013Oct}. Section~\ref{sec4} will then discuss the case of PNM. We will finally provide a short summary of the work presented and draw conclusions in the last section. The explicit expressions for the correlated density-dependent 2B force are given in the Appendix.

\section{Density-dependent two-body interaction at N2LO}
\label{sec2}
A detailed description of the extension of the SCGF method to include 3B forces has been presented in Ref.~\cite{Carbone2013Nov}. Furthermore, in Sec.~II of Ref.~\cite{Carbone2013Oct} we have described the specific extension for the ladder approximation in infinite-nuclear-matter calculations. In this section, we will briefly review the method again, but we will then primarily focus on the construction of the density-dependent force. 

\subsection{3B forces in the SCGF ladder approximation}
\label{sec2a}
The \emph{ladder approximation} is a minimal and consistent SCGF approach to describe a dense medium of strongly-interacting fermions \cite{Dickhoff2008}. In this approach, correlations are taken into account by summing up multiple particle-particle and hole-hole 2B scattering processes in the medium, by means of an in-medium $T$ matrix. Particles are dressed at all stages and off-shell effects are considered via an iterative self-consistent solution of the Dyson equation \cite{Kadanoff1962}. The formalism is thermodynamically consistent, as it fulfils the Hugenholtz-van Hove theorem \cite{Hugenholtz1958}. 

As described in Ref.~\cite{Carbone2013Oct}, we include 3B forces in our infinite matter SCGF calculation via effective operators at the 1B and 2B levels. We deal with a Hamiltonian which is the sum of three terms as follows: 
\beqn
\nn
\h H &=& \sum_{\al} T_\al\, a^\dg_\al a_\al 
+\f 1 4 \sum_{\al\ga\be\de}V_{\al\ga,\be\de}\, a_\al^\dg a_\ga^\dg a_{\de} a_{\be}
\\ 
&&+\frac{1}{36}\sum_{\al\ga\ep\be\de\eta} W_{\al\ga\ep,\be\de\eta}\,
a_\al^\dg a_\ga^\dg a_\ep^\dg a_{\eta} a_{\de} a_{\be}\,.
\label{H}
\enqn
The Greek indices ($\al$, $\be$, $\ga$, etc.) label a complete set of single-particle (SP) states which diagonalize the kinetic single-particle operator, $\h T$. $a^\dg_\al$ and $a_\al$ are creation and annihilation operators for a particle in state $\al$. The matrix elements of $\h T$ are given by $T_\al$. Equivalently, the antisymmetrized matrix elements of the 2B and 3B forces are $V_{\al\ga,\be\de}$ and $W_{\al\ga\ep,\be\de\eta}$.

\begin{figure}[t!]
  \begin{center}
  \subfloat[]{\label{ueffective}\includegraphics[width=0.45\textwidth]{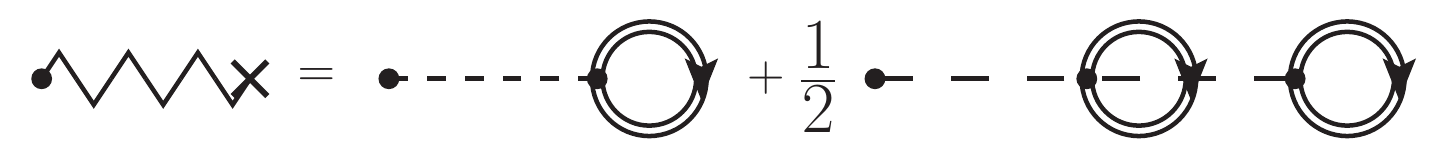}}
  \vspace{.2cm}
  \subfloat[]{\label{veffective}\includegraphics[width=0.45\textwidth]{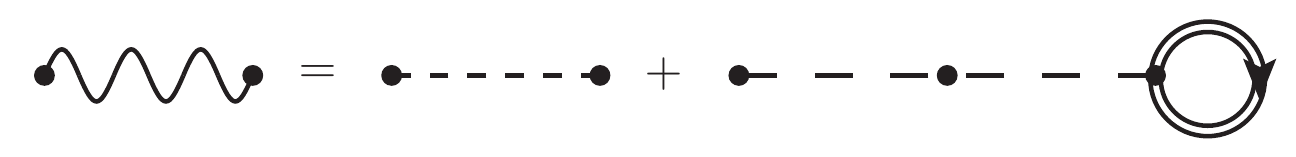}}
  \caption{Diagrammatic representation of the effective 1B and 2B interactions given respectively in Eq.~(\ref{ueff}) and Eq.~(\ref{veff}). The 1B effective operator, \protect\subref{ueffective}, is given by the sum of the 2B interaction (dashed line) contracted with a dressed SP propagator, $G$ (double line with arrow), and the 3B interaction (long-dashed line) contracted with the Hartree-Fock approximation of a dressed 2B propagator $G^{\rm II}$. The correct symmetry factor of 1/2 in the last term is also shown explicitly. The effective 2B operator, \protect\subref{veffective}, is given by the sum of the original 2B interaction (dashed line) and the 3B interaction (long-dashed line) contracted with a dressed SP propagator, $G$.}
  \end{center}
\end{figure}

Starting from the Hamiltonian given in Eq.~(\ref{H}), we can construct effective interactions which include contracted 3B interactions in one- and two-body operators. In doing so, we avoid the task of explicitly calculating 3B interaction-reducible diagrams. In the following calculations, we use the effective operators described in Figs.~\subref*{ueffective} and \subref*{veffective}.
These operators are constructed using an analogous philosophy to the normal ordering of the Hamiltonian given in Eq.~(\ref{H}) \cite{Bogner2010}. Yet the reference state  is now a dressed many-body ground state, which incorporates the correlations of the system. The 1B effective potential in Fig.~\subref*{ueffective} reads
\beqn
\nn
\widetilde U& =&\sum_{\al\be}\Big[
- i\hbar \sum_{\ga\de}V_{\al\ga,\be\de} \, G_{\de \ga}(t-t^+) 
\\ &&  
+ \f{i\hbar}{2} \sum_{\ga\ep\de\eta} 
W_{\al\ga\ep,\be\de\eta}
\,G_{\de\ga}(t-t^+)G_{\eta\ep}(t-t^+)\Big] a_\al^\dg a_{\be}\, ,\quad\,\,
\label{ueff} 
\enqn
and it is the sum of two contributions: a 1B average over the 2B interaction; and a 2B average over the 3B force. We point out that this expression is a first order approximation of the complete 1B effective potential. The full average should include a contraction of the 3B force term with a full 2B propagator \cite{Carbone2013Nov}. However, finite-nuclei Green's function calculations have verified the quality of the first order approximation \cite{Cipollone2013,Barbieri2014}. 

More importantly, and closely related to the main aim of this paper, we note that the 1B and 2B averages in Eq.~(\ref{ueff}) are performed using fully dressed one-body propagators. These originate directly from the correlated reference state which we use in the construction of the effective operators. Similarly, the effective 2B force is given by the expression,
\beqn
\nn
\widetilde V = \f 1 4&&\sum_{\al\ga\be\de} \Big[ V_{\al\ga,\be\de} \\
&&- i\hbar \sum_{\ep\eta}W_{\al\ga\ep,\be\de\eta} \,
G_{\eta\ep}(t-t^+)\Big] a_\al^\dg a_\ga^\dg a_{\de}a_{\be} \,, 
\label{veff}
\enqn
which is obtained from the original 2B interaction plus a 1B average over the 3B force. The operators given in Eq.~(\ref{ueff}) and Eq.~(\ref{veff}), together with a residual 3B irreducible part, define an effective Hamiltonian to be used in the perturbative expansion of the SP propagator \cite{Carbone2013Nov}. As long as \emph{interaction irreducible diagrams} are considered, double counting is avoided in the diagrammatic expansion \footnote{A diagram is \emph{interaction irreducible} if it does not contain any articulation vertex. An interaction vertex is an articulation vertex if it gives rise to two disconnected diagrams when it is cut out of the diagram.}. Moreover, the number of diagrams at each order in the perturbative expansions of the self-energy is substantially reduced  \cite{Carbone2013Nov}. In the self-consistent calculation, the effective operators in Figs.~\subref*{ueffective} and \subref*{veffective} are recalculated at each iterative step, to account for changes in the internal propagators due to the correlations in the system. 

\begin{figure}[t!]
\begin{center}
\includegraphics[width=0.3\textwidth]{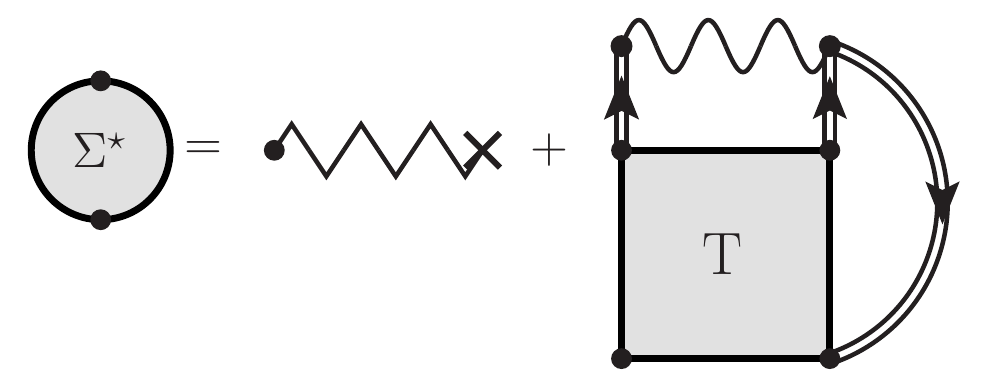}
\caption{Diagrammatic representation of the irreducible self-energy $\Sigma^\star$. The first term is energy independent and it is computed from the effective 1B interaction of Fig.~\protect\subref*{ueffective}. The second term is a dynamical contribution, consisting of excited configurations generated through the 2B effective interactions of Fig.~\protect\subref*{veffective}.}
\label{self_en}
\end{center}
\end{figure}

\begin{figure}[t!]
\begin{center}
\includegraphics[width=0.3\textwidth]{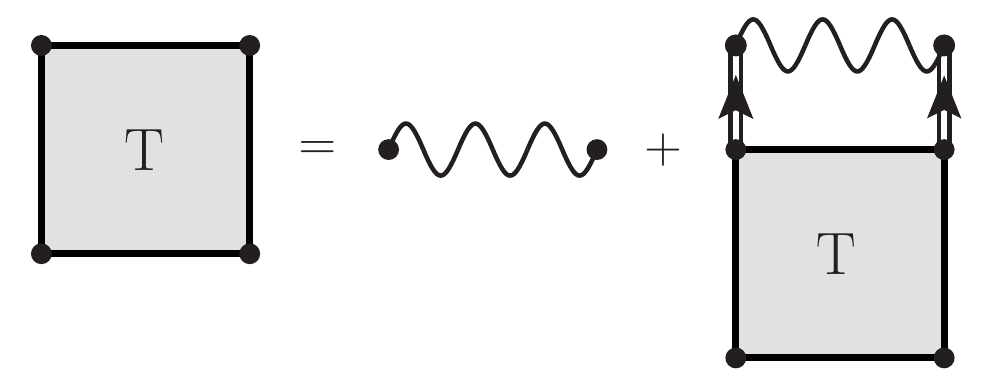}
\caption{Diagrammatic representation of the $T$-matrix interacting vertex function, constructed with the 2B effective interaction of Fig.~\protect\subref*{veffective}.}
\label{tmatrix}
\end{center}
\end{figure}

The proper self-energy $\Sigma^*(\om)$ is implicitly defined by the Dyson equation as follows:
\beq
G_{\al\be}(\om)=G^{(0)}_{\al\be}(\om)+\sum_{\ga \de}G^{(0)}_{\al\ga}(\om)
\Sigma^\star_{\ga\de}(\om)G_{\de\be}(\om)\, ,
\label{Dyson}
\enq
where $G^{(0)}$ ($G$) corresponds to the unperturbed (fully dressed) SP propagator. In the ladder approximation and disregarding three-body irreducible terms, the self-energy is diagrammatically represented by the two contributions depicted in Fig.~\ref{self_en}~\cite{Carbone2013Nov}. 
The first contribution is an energy independent term, which corresponds to the 1B effective potential of Eq.~(\ref{ueff}). In the present approach, this is essentially a correlated Hartree-Fock contribution coming from 2B and 3B forces. The second contribution is a dispersive part which, in the ladder approximation, arises from  the 2B effective potential of Eq.~(\ref{veff}). The $T$-matrix in this contribution is obtained by solving a Lippmann-Schwinger-type equation with the correlated density-dependent 2B effective potential, as shown in Fig.~\ref{tmatrix}.
The $T$-matrix is, in essence, an energy-dependent effective interaction in the medium. After the self-energy has been obtained, one solves the Dyson equation to find the corresponding self-consistent solution of the 1B propagator. We note that the techniques developed to solve the ladder approximation with 2B forces are naturally extended to the 3B sector with this prescription. In particular, there are essentially two additional steps to be considered. First, the Hartree-Fock self-energy is complemented with the 3NF term as in Fig.~\subref*{ueffective}. Second, the two-body potential is replaced by the two-body effective interaction, Fig.~\subref*{veffective}.

The imaginary part of this Green's function provides access to the SP spectral function, $A(p,\om)$, which describes the fragmentation of strength in energy, $\om$, of a state with momentum $p$. The spectral function is then used to compute SP momentum distribution, 
\beq
n(p)=\int\frac{\d\om}{2\pi} A(p,\om)f(\om)\,,
\label{mom}
\enq
where $f(\om)$ corresponds to the Fermi-Dirac distribution function. We note that, in order to prevent the pairing instability below the critical temperature, our calculations are performed at finite temperature \cite{Frick2003}. All the properties presented in Secs.~\ref{sec3} and \ref{sec4} are extrapolated to zero temperature, unless otherwise stated. We use a simple procedure, based on the Sommerfeld expansion \cite{Rios2009Feb}, to find $T=0$ results from low-temperature calculations, usually performed at $T = 5$ MeV. The errors associated to this extrapolation procedure are negligible \cite{Rios2009Feb}. 

Finally, having found the spectral function of the system, we need to compute the bulk thermodynamical properties. The total energy of the system is obtained by means of a modified Galitskii-Migdal-Koltun sum rule, which has been extended to account for 3B forces \cite{Carbone2013Nov}. In the infinite system, this sum rule takes the following form:
\beq
\frac{E}{A}=\frac{\nu}{\rho}\int\frac{\d{\bf p}}{(2\pi)^3}\int\frac{\d\om}{2\pi}\f 1 2 \left\{\frac{p^2}{2m}+
\om\right\}A(p,\om)f(\om) - \f 1 2 \langle W\rangle\,,
\label{sumrule}
\enq
where $\nu$ is the spin-isospin degeneracy of the system and $\rho$ the total density. $\langle W\rangle$ corresponds to the expectation value of the 3NF. In principle, this term would involve a 3B propagator. In the present calculations, however, we evaluate it at the Hartree-Fock level from three, independent but fully dressed, SP propagators. 
\\\\

\subsection{Density-dependent chiral two-body \\ force at N2LO}
\label{sec2b}

\begin{figure}[t!]
  \begin{center}
  \subfloat[]{\label{TPE-3B}\includegraphics[width=0.14\textwidth]{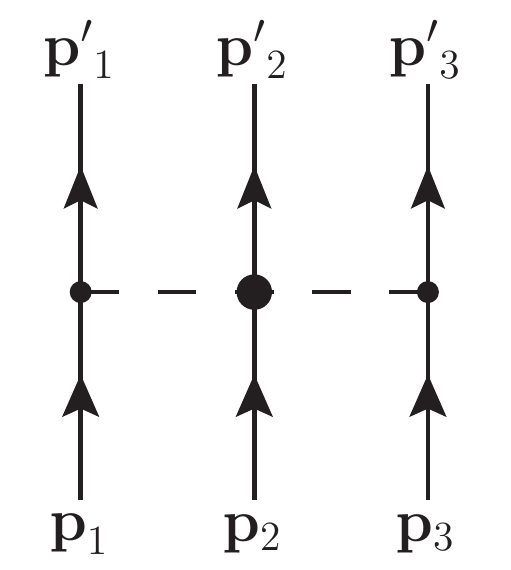}}
  \hspace{.5cm}
  \subfloat[]{\label{OPE-3B}\includegraphics[width=0.13\textwidth]{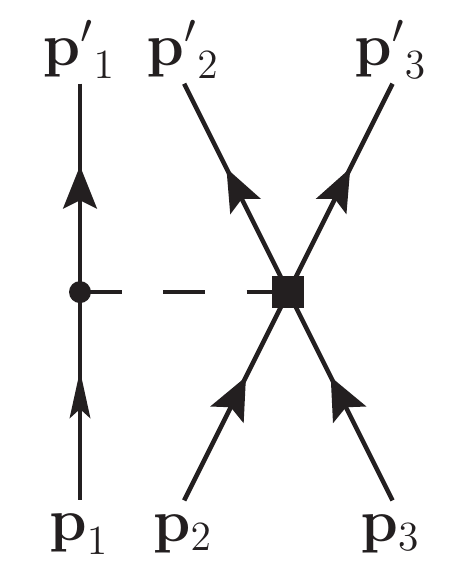}}
  \hspace{.5cm}
  \subfloat[]{\label{cont-3B}\includegraphics[width=0.105\textwidth]{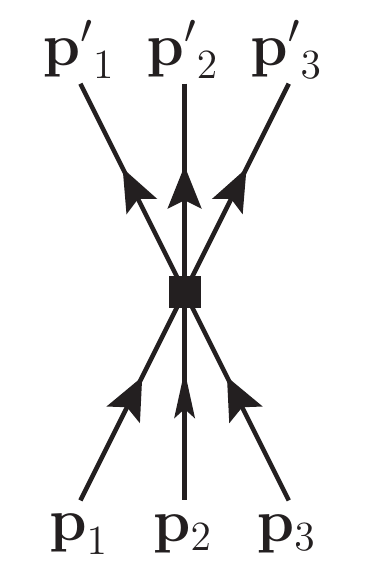}}
  \caption{3NF terms appearing at N2LO in the chiral expansion. Diagram \protect\subref{TPE-3B} corresponds to the TPE term given in Eq.~(\ref{tpe}). Diagram \protect\subref{OPE-3B} is the OPE term of Eq.~(\ref{ope}). Diagram \protect\subref{cont-3B} yields the contact term of Eq.~(\ref{cont}). Dashed lines define pions. Small dots, big dots and squares define the nature of the vertices \cite{Epelbaum2009}.}
  \label{3NF_terms}
  \end{center}
\end{figure}

The 3NFs at N2LO in chiral perturbation theory are given by three terms: a TPE contribution, which corresponds to the Fujita-Miyazawa original $2\pi$ exchange \cite{Fujita1957}; a one-pion-exchange (OPE); and a contact (cont) term \cite{vanKolck1994,Epelbaum2002Dec2}. The diagrammatic representations of these contributions are presented in Fig.~\ref{3NF_terms}.
Their respective analytical expressions are given by the following:
\beqn
W_\mathrm{TPE} &=& \frac{g_A^2}{8F_\pi^4} \sum_{i\neq j\neq k} 
\frac{(\bm{\sigma}_i\cdot{\bf q}_i)(\bm{\sigma}_j\cdot{\bf q}_j)}{({\bf q}_i^2 + M_\pi^2)
({\bf q}_j^2 + M_\pi^2)}
F_{ijk}^{\alpha\beta}\tau_i^{\alpha}\tau_j^{\beta}\,,  \quad
\label{tpe} \\
W_\mathrm{OPE} &=& - \frac{c_D g_A}{8F_\pi^4\Lambda_\chi} \sum_{i\neq j\neq k}
\frac{\bm{\sigma}_j\cdot{\bf q}_j}{{\bf q}_j^2 + M_\pi^2}(\bm{\sigma}_i\cdot{\bf q}_j)
(\bm{\tau}_i\cdot\bm{\tau}_j)\,, 
\label{ope} \\
W_\mathrm{cont} &=& \frac{c_E}{2F_\pi^4\Lambda_\chi} \sum_{j\neq k} 
\bm{\tau}_j \cdot \bm{\tau}_k \,.
\label{cont}
\enqn
In the TPE contribution of Eq.~(\ref{tpe}), $F_{ijk}^{\alpha\beta}$ corresponds to
\beqn
\nn
F_{ijk}^{\alpha\beta}&=&\delta^{\alpha\beta} [-4M_\pi^2c_1+2 c_3{\bf q}_i\cdot{\bf q}_j]  \\
&+&\sum_\gamma c_4\epsilon^{\alpha\beta\gamma}\tau^\gamma_k\bm{\sigma}_k\cdot[{\bf q}_i\times{\bf q}_j]\,.
\label{tpe_tensor}
\enqn
In the previous expressions, ${\bf q}_i={\bf p'}_i- {\bf p}_i$ is the transferred momentum between an incoming and an outgoing nucleon, $i=1,2,3$. ${\bf p}_i$ and ${\bf p'}_i$ are initial and final SP momenta, as depicted in Fig.~\ref{3NF_terms}. $\bm{\tau}_i$ and $\bm{\sigma}_i$ define the corresponding isospin and spin matrices. $g_A=1.29$ is the axial-vector coupling constant, $M_\pi=138.04$ MeV, the average pion mass, $F_\pi=92.4$ MeV, the weak pion decay constant and $\Lambda_\chi=700$ MeV is the chiral symmetry breaking constant, of the order of the $\rho$ meson mass. 

The LECs in  the TPE term, $c_1,\,c_3,\,c_4$, are the same as those appearing in the original 2NF, either the N3LO Entem-Machleidt or the N2LOopt potentials. These are fixed by experimental nucleon-nucleon (NN) phase-shifts and deuteron properties. For Entem-Machleidt, the constants are $c_1=-0.81 \textrm{ GeV}^{-1}, c_3=-3.2 \textrm{ GeV}^{-1}, c_4=5.4 \textrm{ GeV}^{-1}$ \cite{Entem2003}. The derivative-free optimization procedure for N2LOopt, in contrast, provides the values $c_1=-0.92 \textrm{ GeV}^{-1}, c_3=-3.89 \textrm{ GeV}^{-1}, c_4=4.31 \textrm{ GeV}^{-1}$ \cite{Ekstroem2013,Hagen2014}.  

In contrast, the two LECs appearing in the OPE and contact 3NF terms, $c_D$ and $c_E$, are fitted to experimental data in the few-body sector. Their values can be obtained from a variety of methods  \cite{Epelbaum2012}. With the Entem-Machleidt 2NF, we use the LEC values $c_D=-1.11$ and $c_E=-0.66$, taken from fits to ground-state properties of $^3$H and $^4$He \cite{Nogga2006}. This choice is dictated by consistency with the underlying, not renormalized N3LO 2NF. The cutoff on the 3NF included in the regulator function (see Eq.~(\ref{reg_fun})) is chosen to be $\Lambda_\textrm{3NF}=2.5 \textrm{ fm}^{-1}$ in all cases. We note that an error estimate of the underlying nuclear interaction would involve propagating the errors in all these constants to the many-body calculation.

The leading-order 3NF contributions appearing in Eqs.~(\ref{tpe})-(\ref{cont}) are antisymmetrized \cite{Epelbaum2002Dec2}. In other words, the 3B antisymmetrization operator $A_{123}$ is applied to the interaction, where the three-particle antisymmetrization operator reads
\beq
A_{123}=\frac{(1-P_{12})}{2} \frac{(1-P_{13}-P_{23})}{3}\,.
\label{antisymm}
\enq
This is particularly neat, because the antisymmetrization operator splits into a first term that affects particles $1$ and $2$, and a second term involving all three nucleons. $P_{12}$ is the permutation operator of momentum and spin-isospin of particles $1$ and $2$. In spin-isospin space, the operator reads
\beq
P_{12}=\frac{1+\bm{\sigma}_1\cdot\bm{\sigma}_2}{2}\frac{1+\bm{\tau}_1\cdot\bm{\tau}_2}{2}\,.
\label{perm_op}
\enq

The density-dependent 2NF is obtained by tracing over the spin, isospin and momentum of a third, averaged particle. To account for correlations consistently, and following Eq.~(\ref{veff}), the density-dependent effective 2B force is obtained from an average over a dressed SP propagator, involving the correlated momentum distribution $n({\bf p}_3)$. The mathematical expression for the averaged potential is: 
\beqn
\langle {\bf 1' 2'}|\tilde V^\mathrm{3NF}
|{\bf 1 2}\rangle_A =&&
\mathrm{Tr}_{\sigma_3}\mathrm{Tr}_{\tau_3}
\int \frac{{\d}{\bf p}_3}{(2\pi)^3}n({\bf p}_3)
\label{dd3bf}
\\ \nn && 
\langle {\bf 1' 2' 3'}|W
(1-P_{13}-P_{23})|{\bf 1 2 3}\rangle_{A_{12}} \,.
\enqn
For practical reasons, we write $|{\bf 1}\rangle \equiv |{\bf p}_1\bm{\sigma}_1\bm{\tau}_1\rangle$. The subscript $A_{12}$ on the right-hand side matrix element denotes that the three-particle ket is antisymmetrized  with respect to particles $1$ and $2$ only. In other words, the first term of Eq.~(\ref{antisymm}) has already been applied to the three-body ket. The second term of Eq.~(\ref{antisymm}), $(1-P_{13}-P_{23})$, contributes explicitly to the average in Eq.~(\ref{dd3bf}). With this, we ensure that we take into account correctly all possible internal and external permutations. Hence, the final density-dependent 2NF matrix element is properly antisymmetrized at the 2B level. 
 
The momentum integration of Eq.~(\ref{dd3bf}) is regularized, to avoid accessing unphysically large momentum scales. Different regularization functions have been proposed in the past \cite{Hebeler2010Jul,JWHolt2010}. We will explore these differences in the following section. To this end, it is useful to define a general nonlocal regulator that reads:
\beqn
\nn 
f(k,k',p_3)&=&\exp{\left[-\left(\frac{k}{\Lambda_\mathrm{3NF}}\right)^4
-\left(\frac{k'}{\Lambda_\mathrm{3NF}}\right)^4\right]}
\\ && \quad
\exp{\left[-\frac{2}{3}\frac{p_3^2}{\Lambda^4_\mathrm{3NF}}\left(\frac{p_3^2}{3}+(k^2+k'^2)\right)\right]}\,,\,\,\,\quad
\label{reg_fun}
\enqn
where $k=|{\bf k}|=|{\bf p}_1-{\bf p}_2|/2$ and $k'=|{\bf k'}|=|{\bf p}'_1-{\bf p}'_2|/2$ are the moduli of the relative incoming and outgoing momenta. $p_3$ is the modulus of the SP momentum of the averaged particle. $\Lambda_\mathrm{3NF}$ defines the cutoff value applied to the 3NF \cite{Epelbaum2002Dec2}.  We note here that the regulator is symmetric in the interchange of the three particles, and hence it is not affected by the permutations performed in the average.  

The regulator in Eq.~(\ref{reg_fun}) separates naturally into two terms. The first exponential contribution only depends on the external relative incoming and outgoing momenta, $k$ and $k'$.  In contrast, the second term in Eq.~(\ref{reg_fun}) depends on the momentum of the averaged particle and hence acts as an \emph{internal regulator}. In the following, we will refer to \emph{full regulator} when the average is computed using the complete function. Conversely, \emph{external regulator} calculations are obtained when the average is computed using only the first exponential term. The authors of Ref.~\cite{JWHolt2010} use a regulator function which equals only this first term. For this reason, they obtain semi-analytical expressions for the integrals of Eq.~(\ref{dd3bf}). The authors of Ref.~\cite{Hebeler2010Jul} use a full regulator function. There is no physical argument to choose between either regulator. In the following section, we will discuss what differences (if any) emerge when using different regulator functions. 

In this work, the density-dependent two-body force has been derived in the approximation of zero center-of-mass momentum, i.e., ${\bf P}={\bf p}_1+{\bf p}_2=0$. Previous work has shown that this yields very small errors on the bulk properties of infinite matter \cite{Hebeler2010Jul}. In addition, we present  the in-medium density-dependent contributions for diagonal relative momentum matrix elements, $|{\bf k}|=|{\bf k'}|=k$. The in-medium $T$ matrix, however, depends on off-diagonal elements of both the original and the density-dependent 2NF. For the latter, the off-diagonal elements are extrapolated from the diagonal elements following the prescription given in Ref.~\cite{JWHolt2010}. This simplifies the evaluation of the density-dependent terms, and avoids the inclusion of additional operatorial structures in the definition of the general NN potential \cite{Erkelenz1971}, as we will discuss in the appendix.

\begin{figure}[t!]
  \begin{center}
  \subfloat[]{\label{TPE-1}\includegraphics[width=0.13\textwidth]{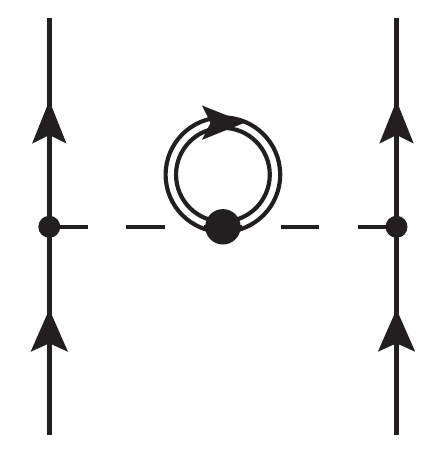}}
  \hspace{1cm}
  \subfloat[]{\label{TPE-2}\includegraphics[width=0.10\textwidth]{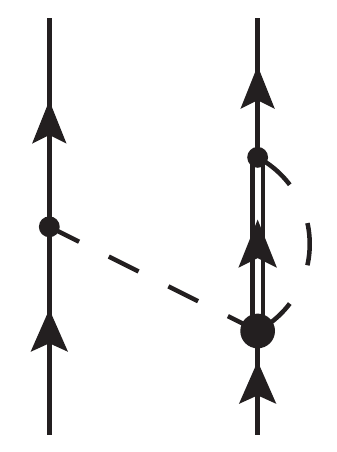}}
  \hspace{1cm}
  \subfloat[]{\label{TPE-3}\includegraphics[width=0.09\textwidth]{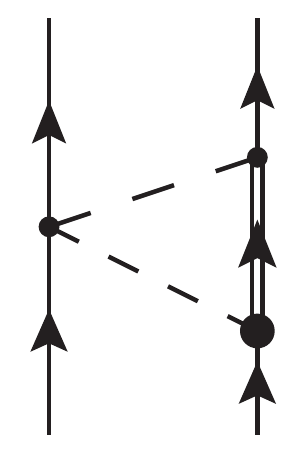}}
  \newline    \vskip .8cm
  \subfloat[]{\label{OPE-1}\includegraphics[width=0.12\textwidth]{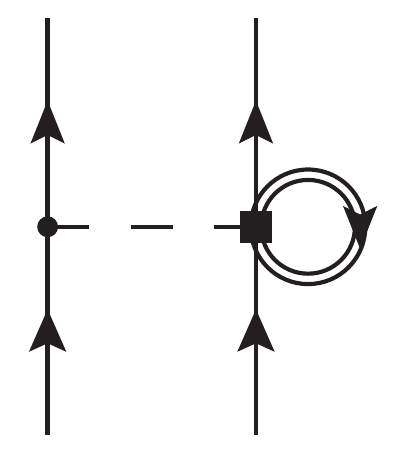}}
  \hspace{1cm}
  \subfloat[]{\label{OPE-2}\includegraphics[width=0.08\textwidth]{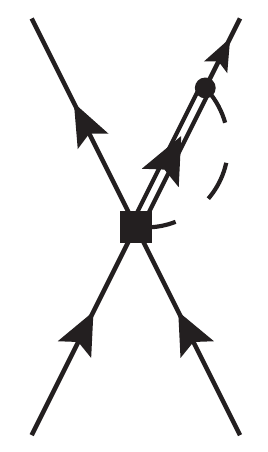}}
  \hspace{1cm}
  \subfloat[]{\label{contact}\includegraphics[width=0.08\textwidth]{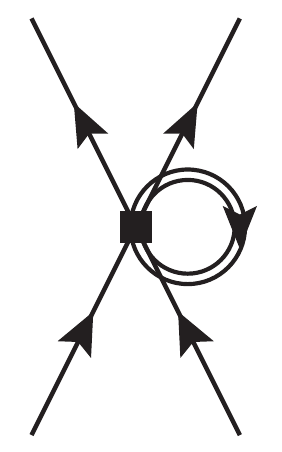}}
  \caption{Six density-dependent contributions arising from contractions of the three 3NF terms appearing at N2LO in the chiral expansion. Dashed lines define pions; double arrowed lines correspond to a dressed single particle propagator. Diagrams \protect\subref{TPE-1}, \protect\subref{TPE-2} and \protect\subref{TPE-3} arise form contraction of the long-range 3NF TPE term given in Eq.~(\ref{tpe}), and correspond respectively to the formal expressions given in the appendix in Eqs.~(\ref{tpe_dd_1}), (\ref{tpe_dd_2}) and (\ref{tpe_dd_3}). Diagrams \protect\subref{OPE-1} and \protect\subref{OPE-2} are obtained from averaging the medium-range 3NF OPE term, Eq.~(\ref{ope}), and correspond respectively to Eq.~(\ref{ope_dd_1}) and Eq.~(\ref{ope_dd_2}). Diagram \protect\subref{contact} is the result of contracting the contact 3NF contribution given in Eq.~(\ref{cont}), and is defined in Eq.~(\ref{cont_dd}). Small and big dots, and squares define the nature of the vertices in the chiral expansion.}
  \label{2NF_dd_terms}
  \end{center}
\end{figure}

If we perform the average of Eq.~(\ref{dd3bf}) for the three 3NF terms at N2LO, Eqs.~(\ref{tpe})-(\ref{cont}), we obtain six density-dependent terms. These are diagrammatically depicted in Fig.~\ref{2NF_dd_terms}. 
The first three terms, Figs.~\subref*{TPE-1}-\subref*{TPE-3}, arise from the contraction of the TPE 3B contribution of Fig.~\subref*{TPE-3B}. Terms given in Figs.~\subref*{OPE-1} and \subref*{OPE-2} arise from the OPE 3B interaction shown in \subref*{OPE-3B}. The last term, Fig.~\subref*{contact}, is the only possible contraction of Fig.~\subref*{cont-3B}. We label these contributions as $\tilde V_\mathrm{TPE-1}^\mathrm{3NF}$, $\tilde V_\mathrm{TPE-2}^\mathrm{3NF}$, $\tilde V_\mathrm{TPE-3}^\mathrm{3NF}$, $\tilde V_\mathrm{OPE-1}^\mathrm{3NF}$, $\tilde V_\mathrm{OPE-2}^\mathrm{3NF}$ and $\tilde V_\mathrm{cont}^\mathrm{3NF}$, respectively. The formal expressions for these six contributions, as well as their derivation, are presented in the appendix. Further details can be found in Ref.~\cite{Carbone2014}. We employ the notation of Ref.~\cite{JWHolt2010}, which we followed to calculate our density-dependent contributions. We note, however, that our expressions differ from those of Ref.~\cite{JWHolt2010} because (a) the regulators are treated differently and (b) correlations are explicitly included in the construction of the in-medium 2B potential.

\subsubsection{Correlation and regulator function effects on the density-dependent 2NF}
\label{sec2b1}

\begin{figure}[t!]
\begin{center}
\subfloat{
\includegraphics[width=0.48\textwidth]{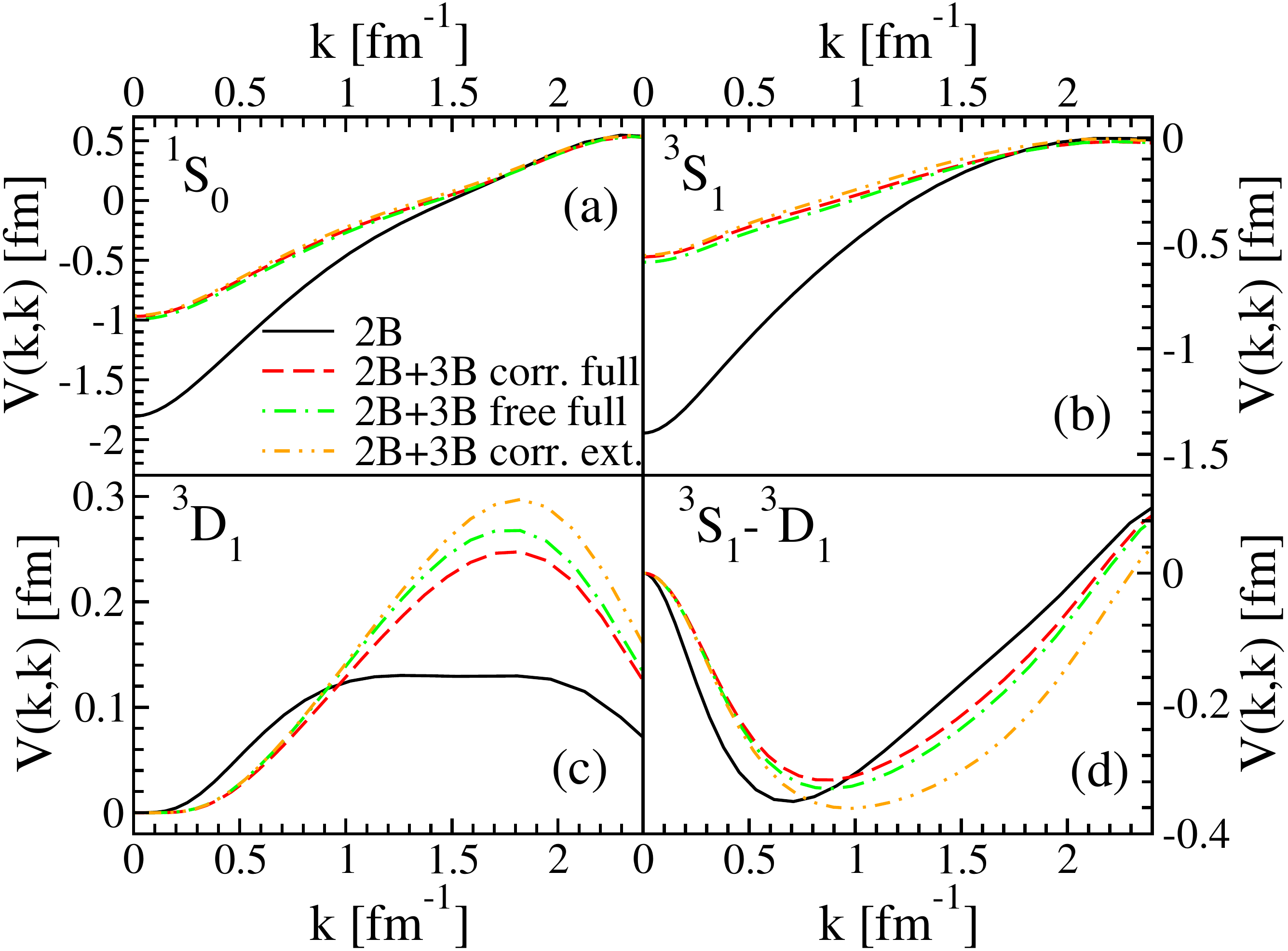}
\label{S_freevsmed_reg2}
}
\hfill 
\subfloat{
\includegraphics[width=0.48\textwidth]{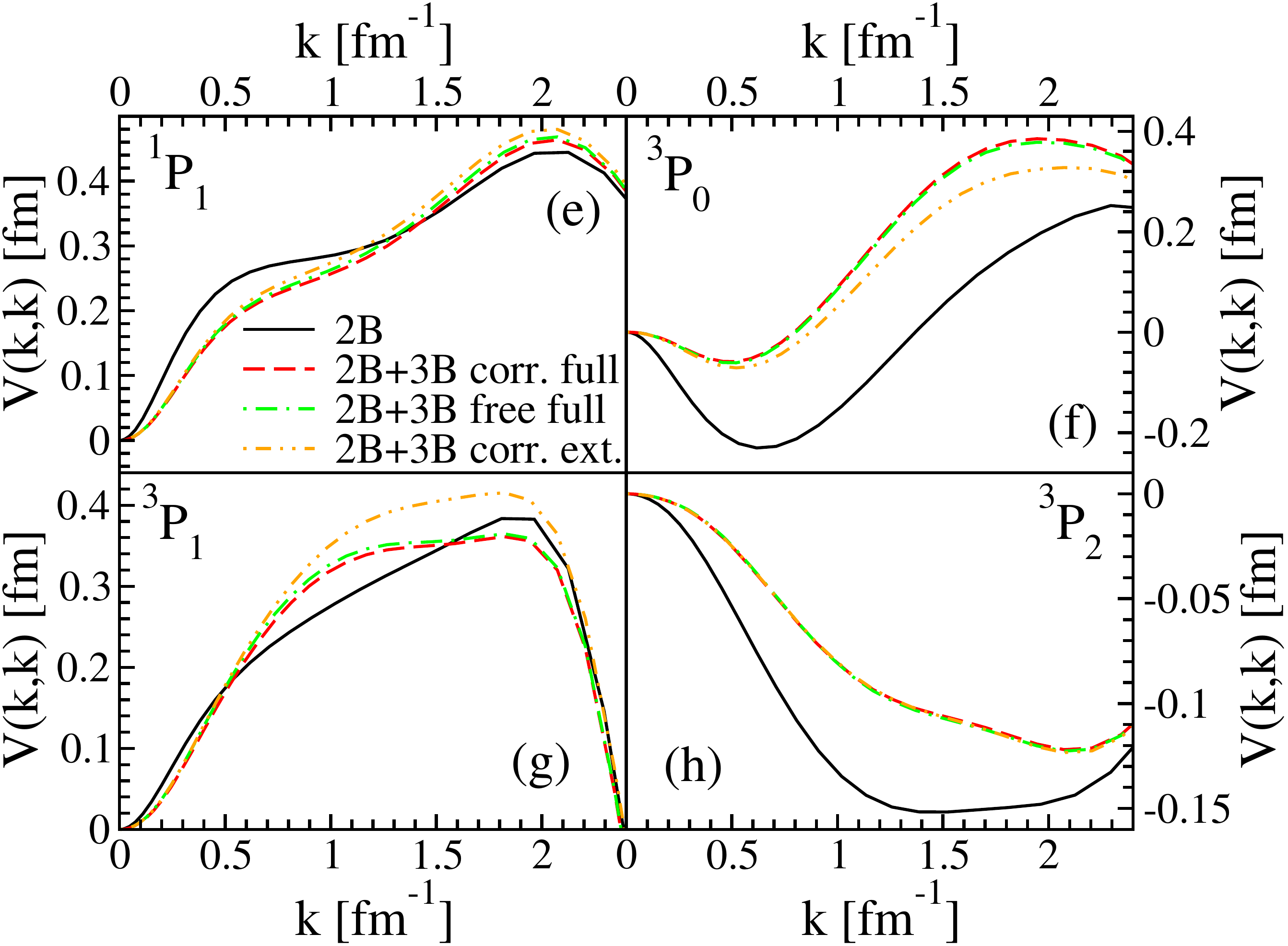}
\label{P_freevsmed_reg2}
}
\caption{(Color online) $S$, $D$, $S-D$ mixing (top panels (a),(b),(c) and (d)) and $P$ (bottom panels (e),(f),(g) and (h)) partial-wave matrix elements of the two-body effective interaction. Black-solid lines depict the bare 2B N3LO potential. Red-dashed, green-dot-dashed and orange-double-dot-dashed curves include the 2B density-dependent force obtained, respectively, in the correlated and free versions with full regulator, and in the correlated version with the external regulator (see explanation in the text). Density-dependent terms are obtained at the empirical saturation density of symmetric nuclear matter, $\rho_0=0.16$ fm$^{-3}$, and for a temperature of $T=5$ MeV.}
\label{partial_waves}
\end{center}
\end{figure}

In this subsection, we analyze the differences which appear in the density-dependent force when performing a correlated or an uncorrelated average in Eq.~(\ref{dd3bf}). We also discuss how the use of a different regulator function affects the integrated results. To start, we plot in Fig.~\ref{partial_waves} the eight lowest partial waves of the 2B force (solid lines).
In addition, we show the corresponding values of the 2B effective potential of Eq.~(\ref{veff}) at the empirical saturation density, $\rho=0.16$ fm$^{-3}$. Curves obtained using both a correlated (dashed line) and a free (dot-dashed line) momentum distribution in Eq.~(\ref{dd3bf}) are presented. The correlated distributions correspond to the self-consistent $n(p)$ obtained via Eq.~(\ref{mom}) at each iteration considering both 2NFs and 3NFs in the calculation. In contrast, a free in-medium propagator corresponds to replacing $n({\bf p}_3)$ in Eq.~(\ref{dd3bf}) by a Fermi-Dirac momentum distribution function, $n(p)=[1+\exp((\vep(p)-\mu)/T)]^{-1}$, where the SP spectra $\vep(p)$ and the chemical potential $\mu$ are consistently calculated in the iterative procedure. The temperature $T$ is equal to 5 MeV. These two curves are obtained using the full regulator of Eq.~(\ref{reg_fun}). Furthermore, we also present in Fig.~\ref{partial_waves} the results obtained with an external regulator (double-dot-dashed line). 

Focusing on the $S$ partial waves in Figs.~\ref{partial_waves}(a) and \ref{partial_waves}(b), we find that the density-dependent 2NF is less attractive than the original N3LO 2B force for all momenta. An analysis of each single density-dependent contribution shows that this modification is mainly caused by two terms: $\tilde V_\mathrm{TPE-3}^\mathrm{3NF}$, which includes medium effects in the TPE 2B term, as depicted in Fig.~\subref*{TPE-3}, and the contact term, $\tilde V_\mathrm{cont}^\mathrm{3NF}$, shown in Fig.~\subref*{contact} \cite{Carbone2014}. The repulsion provided by the density-dependent force is as large as $50\%$ of the value of the bare 2NF at zero momentum. If we now focus on the $D$-wave, in Fig.~\ref{partial_waves}(c),  we observe that the inclusion of the contracted 3NFs provides some small attraction at low momenta, $\Delta V \approx 0.02$ fm, which then evolves in a strong repulsion at intermediate momenta, $\approx 0.1$ fm. In this case, since the contact term is inactive, it is the $\tilde V_\mathrm{TPE-3}^\mathrm{3NF}$ contribution that plays the larger role in providing repulsion \cite{Carbone2014}. The behavior of the mixing $S-D$ wave is reversed with respect to the $D$ wave. When including 3NFs, a small repulsion of the order of $\sim0.02$ fm is found at momenta up to $1$ fm$^{-1}$. At higher momenta, the mixed matrix elements from the density-dependent force are more attractive than the original 2NF. This result is a consequence of the combined effect of all in-medium terms in this partial wave \cite{Carbone2014}.

While a complete characterization of the energy in terms of partial waves is difficult in the SCGF approach, we suspect that the modifications on the $S$ and $D$ waves are the dominant ones for the saturation mechanism associated to chiral 3NFs. We have checked that the repulsive change which characterizes these partial waves grows with the density. This shifts the total energy of the system to more repulsive values with increasing density, providing a mechanism for nuclear matter saturation. 
 
Regarding the use of a dressed or a free propagator, we observe that the largest differences appear in the $D$ and $S-D$ mixing waves. We find that when undressing the propagator, the absolute value of these potential matrix elements increases at intermediate momenta by about $\sim 0.04$ fm. For the $S$ wave, a similar modification appears of the order of $0.03$ fm at small relative momenta. An analysis of the six density-dependent components shows that the major difference between dressed and undressed averages arises from the $\tilde V_\mathrm{TPE-3}^\mathrm{3NF}$ contribution. This term increases in absolute value when going from the correlated to the free average. The structure of this term, Eq.~(\ref{tpe_dd_3}), is rather cumbersome, so it is difficult to ascribe this effect to a single cause. Our analysis suggests, however, that the differences are a consequence of the availability of momenta in the integration of Eq.~(\ref{dd3bf}) \cite{Carbone2014}. 

A further justification of this hypothesis follows from the differences among the results  obtained with or without the internal regulator. The results obtained when momenta are not internally regulated, labeled ``2B+3B corr. ext." in Fig.~\ref{partial_waves}, suggest that the absolute values of matrix elements increase with respect to fully regulated averages. Because the internal momentum is not regulated, the correlated external average has access to larger internal integrated momenta with respect to the other averaging procedures. For this reason, the corresponding matrix elements increase in size.

Similar conclusions hold for higher partial waves. In Figs.~\ref{partial_waves}(e)-\ref{partial_waves}(h) we show the corresponding results for the density-dependent matrix elements of the $P$ waves. In this case, the largest modifications are associated to the three contributions derived from averaging the TPE 3NF, $\tilde V_\mathrm{TPE-1}^\mathrm{3NF}$, $\tilde V_\mathrm{TPE-2}^\mathrm{3NF}$  and $\tilde V_\mathrm{TPE-3}^\mathrm{3NF}$ \cite{Carbone2014}. In the $^3P_0$ and $^3P_2$ waves, density-dependent 2NF are slightly more repulsive at all momenta than the corresponding bare 2B N3LO potential. In the $^3P_0$ wave, the matrix elements increase by about $\approx 0.1$ fm at intermediate momentum. For the $^3P_2$ wave, the increase is slightly smaller, reaching a maximum of $\approx 0.05$ fm at  intermediate momenta. In the $^1P_1$ partial wave, density-dependent 2NF are slightly more attractive (repulsive) than the bare interaction below (above) $1.2$ fm$^{-1}$. The behavior is reversed for the $^3P_1$ partial wave. In both cases, the modification is never higher than $\sim0.04$ fm in absolute value. 

In the $P$ partial waves, results obtained performing the different averages follow the trend already observed for the $D$ and $S-D$ waves. The difference between the different averages, however, are quantitatively smaller in these partial waves. Compared to the full correlated average, an undressed propagator in the $^1P_1$ and $^3P_1$ waves adds a repulsion for all momenta of the order of $\approx 0.01$ fm. In contrast, in the $^3P_0$ wave the effect has the opposite sign. In agreement with the previously observed results and with our hypothesis, the differences are larger when comparing full and external regulator averaging procedures. 

As a concluding remark to this section, we want to stress our hypothesis once again. The comparison of both uncorrelated and correlated but external density dependent matrix elements to fully regulated and correlated ones indicates that access to more high momenta in the averaging procedure yield larger matrix elements in absolute value. In other words, attractive (repulsive) density-dependent forces become more attractive (repulsive) when going from the correlated and fully regulated average to the free or externally regulated matrix elements. In other words, it is the quantity $n(p_3) f(k,k',p_3)$ in the integration of Eq.~(\ref{dd3bf}) which plays the leading role in the modifications of the partial waves. In the next section, we will investigate in detail the many-body results arising from these different averaging procedures.

\section{Nuclear Matter}
\label{sec3}

\begin{figure}[t!]
\begin{center}
\includegraphics[width=0.47\textwidth]{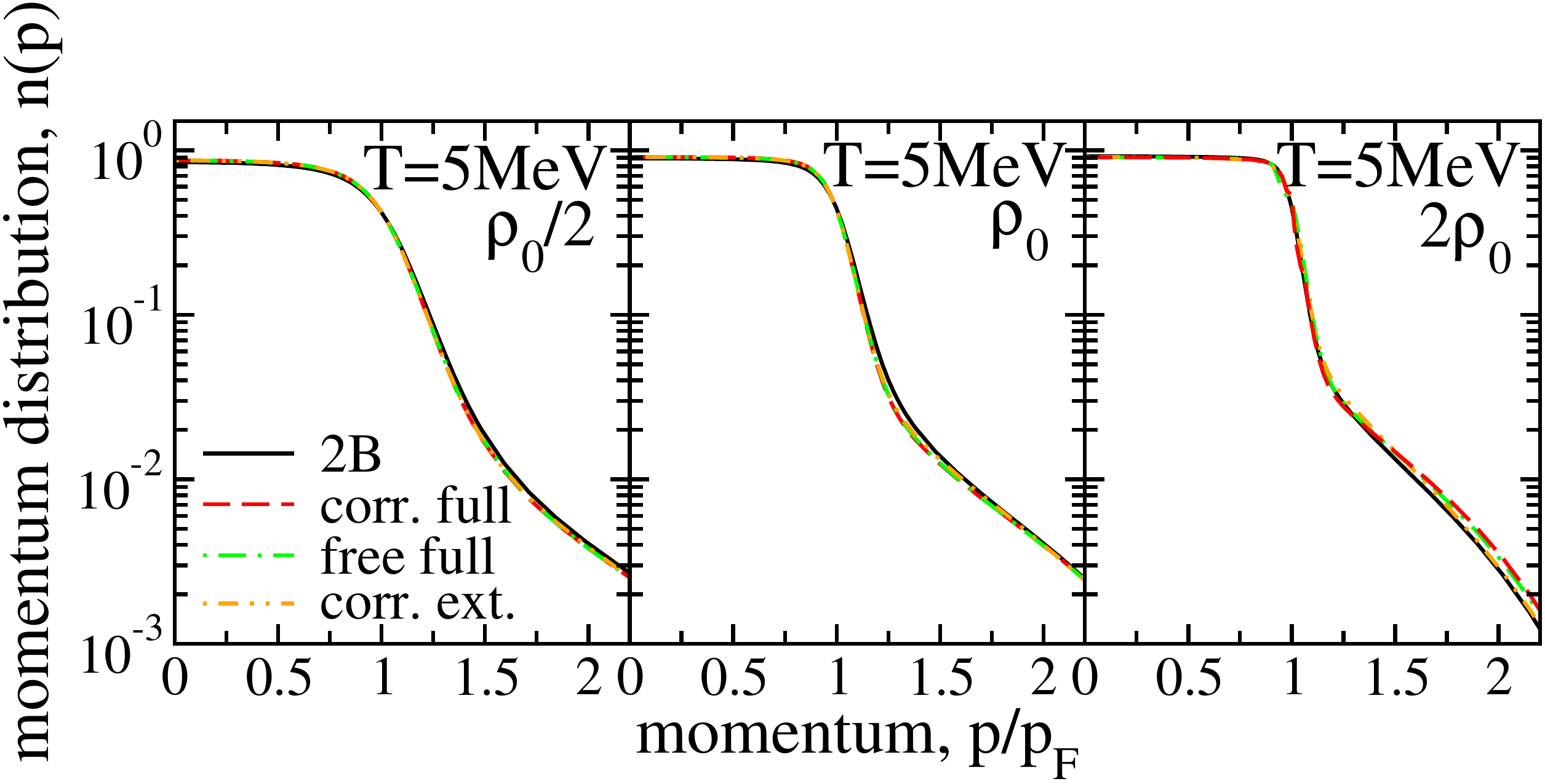}
\caption{(Color online) Momentum distribution in SNM using 2NF at N3LO (black-solid lines) and including the density-dependent 2NF via different averaging procedures. Red-dashed, green-dot-dashed and orange-double-dot-dashed curves correspond to results with 2B density-dependent forces obtained, respectively, with a correlated and uncorrelated distribution with full regulator, and with the correlated version with the external regulator. Calculations are performed at $\rho = 0.08, 0.16$ and $0.32$ fm$^{-3}$ in panels going from left to right. }
\label{momdis_comp}
\end{center}
\end{figure}

The spectral function gives direct access to the SP momentum distribution function, see Eq.~(\ref{mom}). In Fig.~\ref{momdis_comp}, we show the self-consistent solution for $n(p)$, obtained via Eq.~(\ref{mom}), in the case of SNM at  $T=5$ MeV for three densities, $\rho= 0.08$ fm$^{-3}$ (left panel), $0.16$ fm$^{-3}$ (central panel) and $0.32$ fm$^{-3}$ (right panel). 
The usual features of the momentum distribution of a correlated fermionic system are observed. There is a small depletion of population below the Fermi surface, $p_F$. A sharp decrease is observed around $p \approx p_F$, an effect which is softened in our calculations due to finite temperature. Because the momentum distribution is normalized, the strength of the depleted states below $p_F$ is promoted to high momenta above the Fermi surface. There, we observe a steady, exponential decrease of the distribution with momentum. 

We compare results obtained without and with the density-dependent 2NF, calculated using the different averaging procedures presented in the previous section. We recall that the $n(p)$ plotted in Fig.~\ref{momdis_comp} is consistently used in the evaluation of Eq.~(\ref{dd3bf}) when the correlated version of the density-dependent force is performed. The effect of the 3NFs on the momentum distribution is relatively small at all densities. At and below saturation, the difference among calculations with and without 3NFs is negligible. Thanks to the logarithmic scale in Fig.~\ref{momdis_comp}, we appreciate a modification in $n(p)$  for momenta higher than the Fermi momentum at $2\rho_0$. Calculations with the density dependent 2NF have larger high momentum components. This increase at high momenta is a consequence of the additional correlations induced by 3NFs. 

\begin{figure}[t]
\begin{center}
\includegraphics[width=0.45\textwidth]{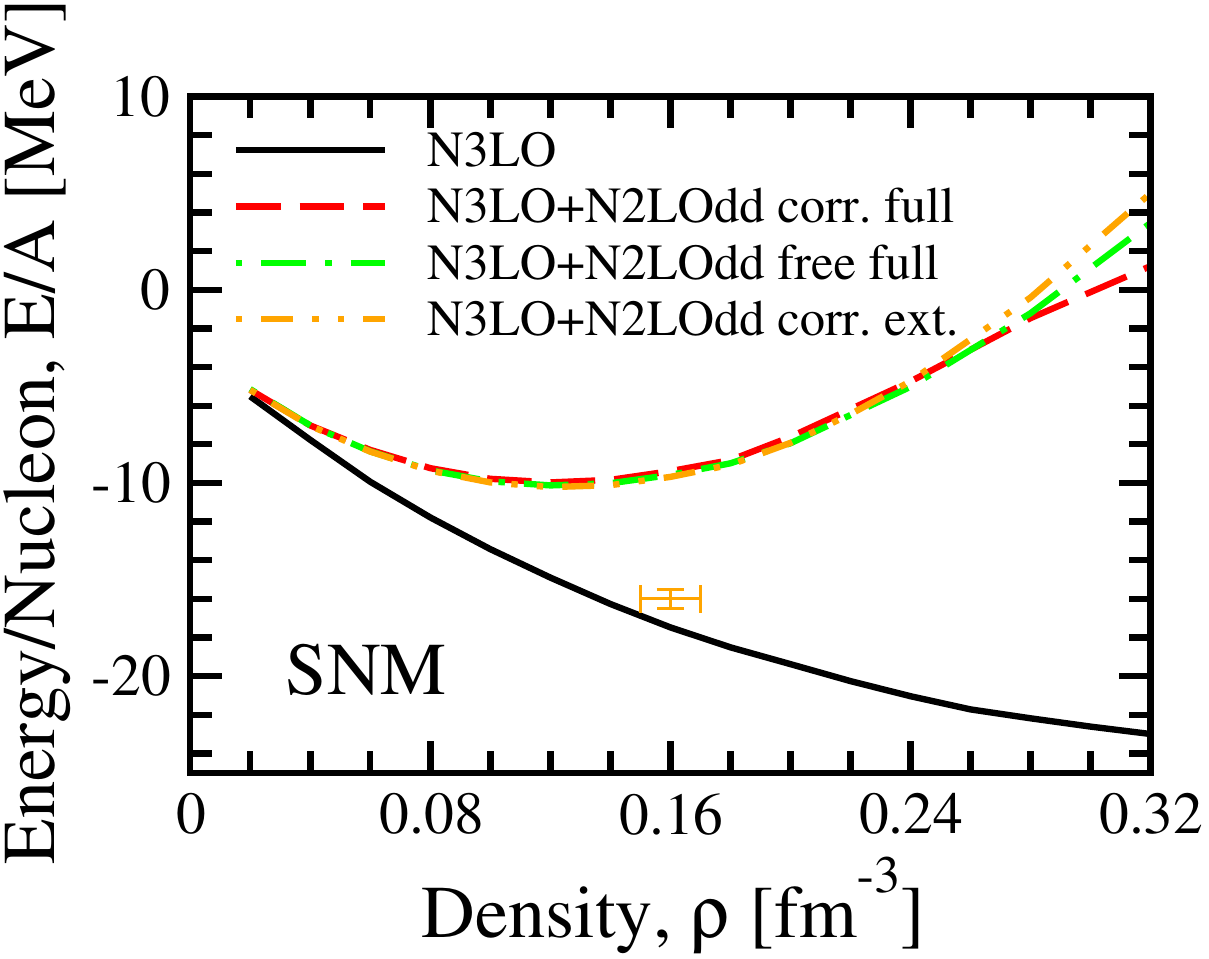}
\caption{(Color online) SNM energy per nucleon as a function of density at $T=0$ MeV. The black-solid line shows the 2B N3LO calculation.  Red-dashed, green-dot-dashed and orange-double-dot-dashed curves correspond to results with 2B density-dependent forces obtained, respectively, with a correlated and uncorrelated distribution with full regulator, and with the correlated version with the external regulator. The orange cross represents the empirical saturation point.}
\label{snm_comp_Nogga}
\end{center}
\end{figure}

Figure \ref{momdis_comp} suggests that the variations of the momentum distribution function due to the different averaging procedures in the construction of the density-dependent force are very small. Concentrating on the low momentum region,  we find that the absolute value of the depletion changes by less than $1\%$. The largest modifications are observed at high densities. At density $2\rho_0$, a small spread in the different curves is observed at high momentum. We observe that the largest high-momentum population is induced by the correlated average with full regulator. Because of these large high-momentum components, we expect a higher total kinetic energy with this average. Nonetheless, the total energy with the correlated average with a full regulator is the less repulsive of all constructions of the averaged force (see Fig.~\ref{snm_comp_Nogga}). This indicates that the potential energy is also sensitive to the averaging procedure. Following our arguments from the previous section, the more regulated momenta should lead to a less repulsive potential and hence to a smaller potential energy.

At this point, we stress that, in spite of the cutoff in both the 2NF and 3NF, a substantial population of high-momentum components is found. Similarly,  the spectral functions display qualitatively important tails at high energies \cite{Carbone2014}. Traditional microscopic 2NF would yield even larger high-momentum components \cite{Rios2014}. Our results highlight the importance of considering such effects in many-body calculations, even with relatively soft interactions like chiral forces. In particular, let us stress that even when using these soft interactions, the low-momentum SP properties are affected by correlations. A depletion of around $\approx 10 \%$ below $p_F$ is typical in our results. 

We now discuss results for the total energy of symmetric nuclear matter obtained using the extended SCGF formalism which includes 3B forces \cite{Carbone2013Oct,Carbone2013Nov}. As already explained, we include 3NF at N2LO in the chiral expansion, constructing a density-dependent 2B potential via an average over the third particle. In the calculation, we include partial waves up to $J=4$ ($J=8$) in the dispersive (Hartree-Fock) contributions. The total energy is computed via the modified Galitskii-Migdal-Koltun sum rule defined in Eq.~(\ref{sumrule}), where the 3B expectation value is evaluated only at the HF level. To perform the expectation value of $\langle W\rangle$ we exploit the 3B term of the 1B effective interaction [see the second term on the right-hand side of Eq.~(\ref{ueff})]. Namely we integrate this quantity over the correlated momentum distribution and multiply by a factor 1/3 to account for the symmetry properties of $\langle W\rangle$ \cite{Carbone2014}. When no 3B forces are included, the standard GMK sum rule is applied \cite{Dickhoff2004}. As already mentioned in Sec.~\ref{sec2a}, we extrapolate $T=0$ results from finite temperature data by relying on the Sommerfeld expansion \cite{Rios2009Feb,Carbone2014}. At low temperatures, the Sommerfeld expansion indicates that the effect of temperature is quadratic for both the energy and the free energy, but with opposite sign \citep{Rios2009Feb}. In other words, the finite temperature energy and free energy read, respectively, $e(T) \sim e_0+a T^2$ and $f(T)  \sim e_0-a T^2$. Consequently, the semisum of both quantities calculated at $T=5$ MeV is an estimate of the zero-temperature energy, $\tfrac{e(T)+f(T)}{2} \approx e_0$. 

In Fig.~\ref{snm_comp_Nogga}, we show the curves of the energy per nucleon obtained for SNM at $T=0$ MeV.
Again, we compare results obtained with a 2NF only (solid lines) to results including a density-dependent 2NF obtained from a variety of averaging procedures. The effect of 3NFs is significant. As expected from the partial-wave analysis, 3B forces induce a substantial repulsion in the energy values. The amount of repulsion increases with density and it is the main cause of nuclear-matter saturation. When including 3NFs, we obtain a qualitatively good saturation density of $\rho \sim 0.14$  fm$^{-3}$ and an underbound saturation energy of $\frac{E}{A} \sim-10$ MeV. In contrast, the 2B calculation, obtained with the N3LO potential of Ref.~\cite{Entem2003}, saturates at densities beyond the range of the figure. To be more precise, saturation is observed at $\rho=0.40$ fm$^{-3}$ and $\frac{E}{A} \sim-24$ MeV. This extremely high saturation density is remedied by the density-dependent 2NF, which shifts the minima of the energy to densities close to the empirical value. 

In agreement with the momentum distribution results of Fig.~\ref{momdis_comp}, the different averaging procedures play a minor role in the total energy. At saturation density, the curves obtained including the contracted 3NFs show minute differences of less than $\sim 0.1$ MeV. Within these differences, the trends are as expected from the partial-wave analysis presented in Sec.~\ref{sec2b}. Specifically, at the saturation density $\rho=0.14$ fm$^{-3}$, matter is less bound when using the correlated average with the full regulator (dashed line). The free average with full regulator (dot-dashed line) is slightly more bound, and the correlated average with the external regulator (double-dot-dashed line) provides the highest binding energy. As already stressed, this shift is extremely small, less than 2\% going from the most attractive to the most repulsive result. 

A larger spread is observed in the energy curves at high densities. Compared to what is observed at saturation density, at $\rho = 0.32$ fm$^{-3}$ the behavior for the curves is reversed. The correlated average with full regulator leads now to the lowest energy value. The correlated average with external regulator presents the more repulsive values, whereas the free average with full regulator is about halfway between the two. In absolute values, the effect of the different averages in the total energy is the same, both at $\rho_0$ and at 2$\rho_0$. In fact, this is a consequence of the arguments we put forward in Sec.~\ref{sec2b}. The higher the availability of momentum states in the averaging integration, the stronger the effect of the 3NF, whether in attraction or in repulsion. 

In any case, the differences between the various average procedures are small, within $3$ MeV of each other up to twice saturation. This result is remarkable in that it validates previous calculations performed with free propagators in the internal average \cite{Hebeler2010Jul,Hebeler2011,JWHolt2010}. In other words, calculations for both micro- and macroscopic properties of SNM with chiral forces are rather insensitive to the regulators that are used. In addition, and rather surprisingly, there is also a tiny dependence on the use of correlations in the internal averaging procedure. This suggests that the effect of single-particle correlations in the construction of the density-dependent force is well under control.

In this context, one can only hypothesize as to why the saturation point that we obtain is still somewhat far away from the empirical point. There are at least two factors that influence this result. First, the underlying interaction is not completely set by the central values of the LECs. Variation on the LECs and the regulator functions will transform each of the single lines of Fig.~\ref{snm_comp_Nogga} into bands. Because of the large number of LECs and regulators acting in SNM, the propagation of errors from the interaction into the many-body calculations can be difficult to quantify \cite{Hebeler2011,Carbone2013Oct}. We will attempt a first-order exploration of these effects in PNM in the following section. Second, the averaging procedure can still be improved. The two-body average in the one-body effective interaction should be performed with a fully correlated two-body propagator. Similarly, the extended sum rule is now computed with a Hartree-Fock 3B expectation value. Alternatives to this sum rule, involving two-body Green's functions, will be explored in the future \cite{Carbone2013Nov}.

Before closing this section, we must point out that some computational difficulties are encountered for SNM at double saturation density. When using the correlated average with a full regulator, numerical instabilities were found in the iterative, self-consistent procedure. To overcome these complications, we rely on the similarity of the momentum distributions obtained with the different averaging procedures (see Fig.~\ref{momdis_comp}). Instead of using a self-consistent $n(p_3)$ at each iteration, we have set the momentum distribution to equal that of the converged result with the uncorrelated average. This momentum distribution is kept fixed at each iteration step. The convergence pattern is substantially improved with this procedure. We note that the momentum distributions obtained in this approach are still self-consistent, in that they enter all the iterative procedure except for the determination of the density-dependent matrix elements. The existence of these instabilities points to a strong non-linearity in the resolution of the ladder approximation with 3NFs.

\section{Neutron Matter}
\label{sec4}

Using the extended SCGF method to include 3B forces, we compute in this section the bulk properties of PNM. This is particularly interesting in the context of nuclear chiral interactions, because up to N3LO in the EFT expansion all many-body forces among neutrons are predicted. In other words, no coupling constants other than those present in the 2B sector ($c_1$ and $c_3$) need to be adjusted. The 3NF terms proportional to $c_4$, $c_D$ and $c_E$ vanish at N2LO \cite{Tolos2008,Hebeler2010Jul}. Recently, studies for PNM have been extended to include full N3LO chiral interactions in perturbative many-body calculations \cite{Tews2013,Krueger2013}.

In PNM, as for the symmetric nuclear matter case, we perform calculations complementing the N3LO 2NF of Ref.~\cite{Entem2003} with the density-dependent force computed in neutron matter. We will also show results obtained using the optimized 2B N2LO chiral force of Ref.~\cite{Ekstroem2013}, which will therefore be consistent in the order of the chiral expansion. In the appendix, we discuss in detail the modifications that are needed to compute density-dependent forces in PNM. 

\begin{figure}[t]
\begin{center}
\includegraphics[width=0.45\textwidth]{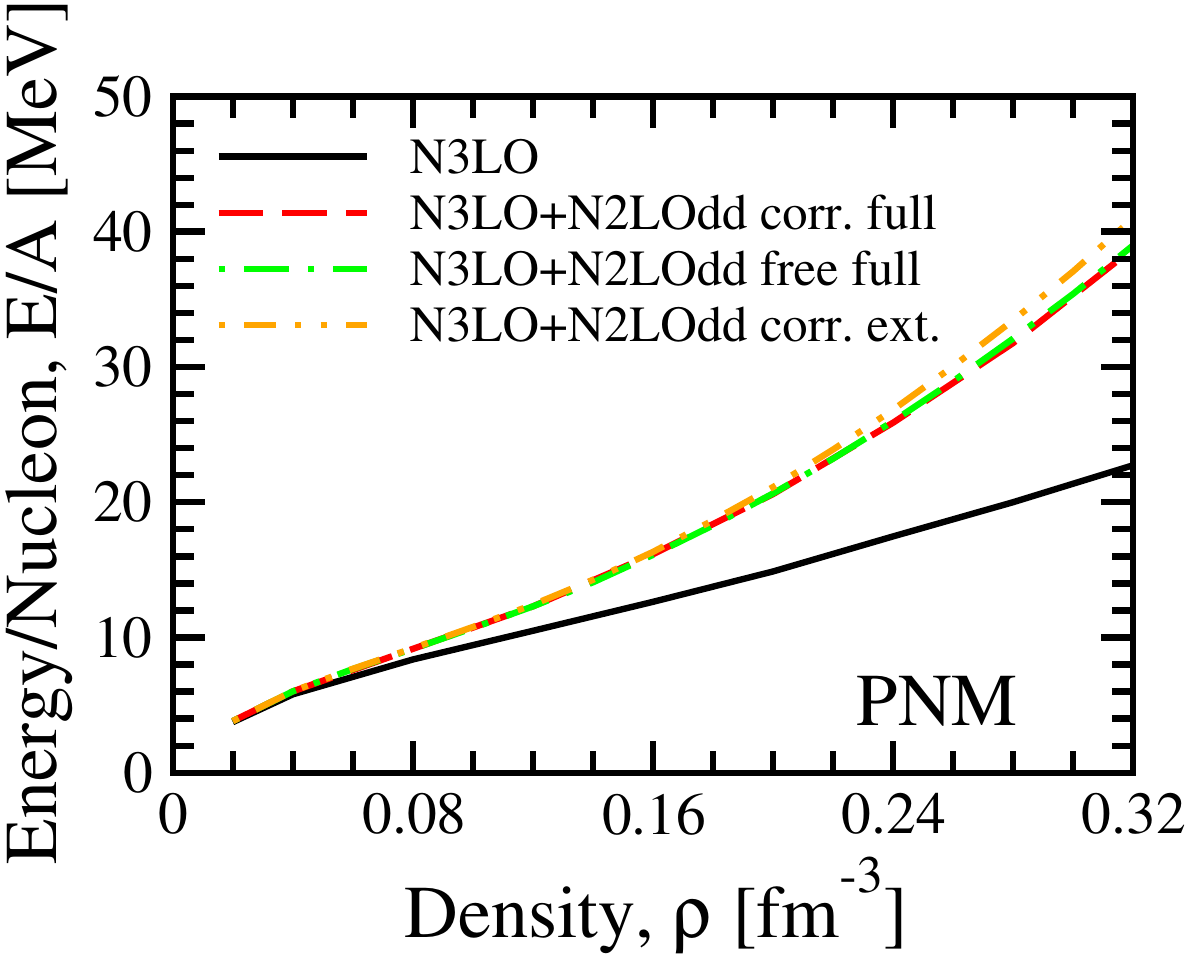}
\caption{(Color online) PNM energy per nucleon as a function of density at $T=0$ MeV. The black-solid line shows the 2B N3LO calculation.  Red-dashed, green-dot-dashed and orange-double-dot-dashed curves correspond to results with 2B density-dependent forces obtained, respectively, with a correlated and uncorrelated distribution with full regulator, and with the correlated version with the external regulator. }
\label{pnm_comp}
\end{center}
\end{figure}

We present in Fig.~\ref{pnm_comp} the density dependence for the total energy per nucleon in PNM extrapolated at zero temperature. 
The density-dependent 2NFs bring in repulsion for all densities. Similarly to SNM, this repulsion rises with increasing density. A stronger density dependence will result in a stiffer equation of state for PNM. This is necessary in order to yield neutron stars with large enough radii and masses \cite{Hebeler2013Jul}. As a matter of fact, the recent observation of pulsars with masses $>2 M_\odot$ have ruled out a variety of microscopic descriptions which lead to softer equations of state \cite{Demorest2010,*Antoniadis2013}.

In the 2B only case plotted in Fig.~\ref{pnm_comp}, the total energy per neutron is around $\sim13$ MeV at the empirical saturation density, $\rho_0=0.16$ fm$^{-3}$. The energy increases around $\sim10$ MeV up to double saturation density, $2\rho_0$. The inclusion of 3NFs contributes to increase the energy in the entire density range. The repulsion caused by the density-dependent 2NF goes from less than $1$ MeV at half saturation to $\approx 4$ MeV at saturation density. As already mentioned, the repulsive effect increases with density and boosts up to $\sim15$ MeV at double saturation density. We point out that, thanks to the nonperturbative nature of our calculation, we are not bound to low-density regions. With a single many-body technique, we can access the high density range. However, care must be taken into account at higher densities due to the breaking down of the chiral expansion. To avoid this inconvenience, previous calculations, at those higher densities, relied on extrapolations of the energy from low density regions, by means of polytropic expansions \cite{Hebeler2013Jul}. 

We also present in Fig.~\ref{pnm_comp} the variation in the total energy curve due to the different averaging procedures in the construction of the 2B density-dependent force. As observed for SNM, the difference is negligible for low densities, and becomes appreciable only well above saturation. At $2\rho_0$, the three different averages differ by about $\sim 2$ MeV. Over a total value of $40$ MeV, this represents a small $5 \%$ variation, in agreement with the differences in SNM of Fig.~\ref{snm_comp_Nogga}. Again, we stress that the curves overlap each other, hence there is a substantial independence of the energy per nucleon on the averaging procedure. 

In SNM, the theoretical uncertainties are dominated by errors on several LECs, including $c_D$ and $c_E$. In contrast, in PNM the uncertainties associated to LECs are only due to variations in the determination of $c_1$ and $c_3$ \cite{Hebeler2010Jul,Hebeler2013Jul}. These LEC relate NN, $\pi$N and 3N interactions  in the chiral expansion. Their determination from $\pi$N scattering is, within uncertainties, in agreement with the NN scattering extracted values. An analysis of these uncertainties leads to the values $c_1=-(0.7 \mbox{-} 1.4)$ GeV$^{-1}$ and $c_3=-(3.2 \mbox{-} 5.7)$ GeV$^{-1}$, as described in Ref.~\cite{Hebeler2010Jul}. As an initial estimate of error propagation associated to these uncertainties, one can perform many-body calculation with different values of these constants. Taking the largest differences, one obtains an error band for the energy per particle as a function of density which reflects the uncertainty in the underlying LECs. Note that in the cases studied here the more repulsive (attractive) values of the energy are always associated to the lower (upper) bounds for the LECs. No numerical instabilities have been observed in PNM.

\begin{figure}[t]
\begin{center}
\includegraphics[width=0.45\textwidth]{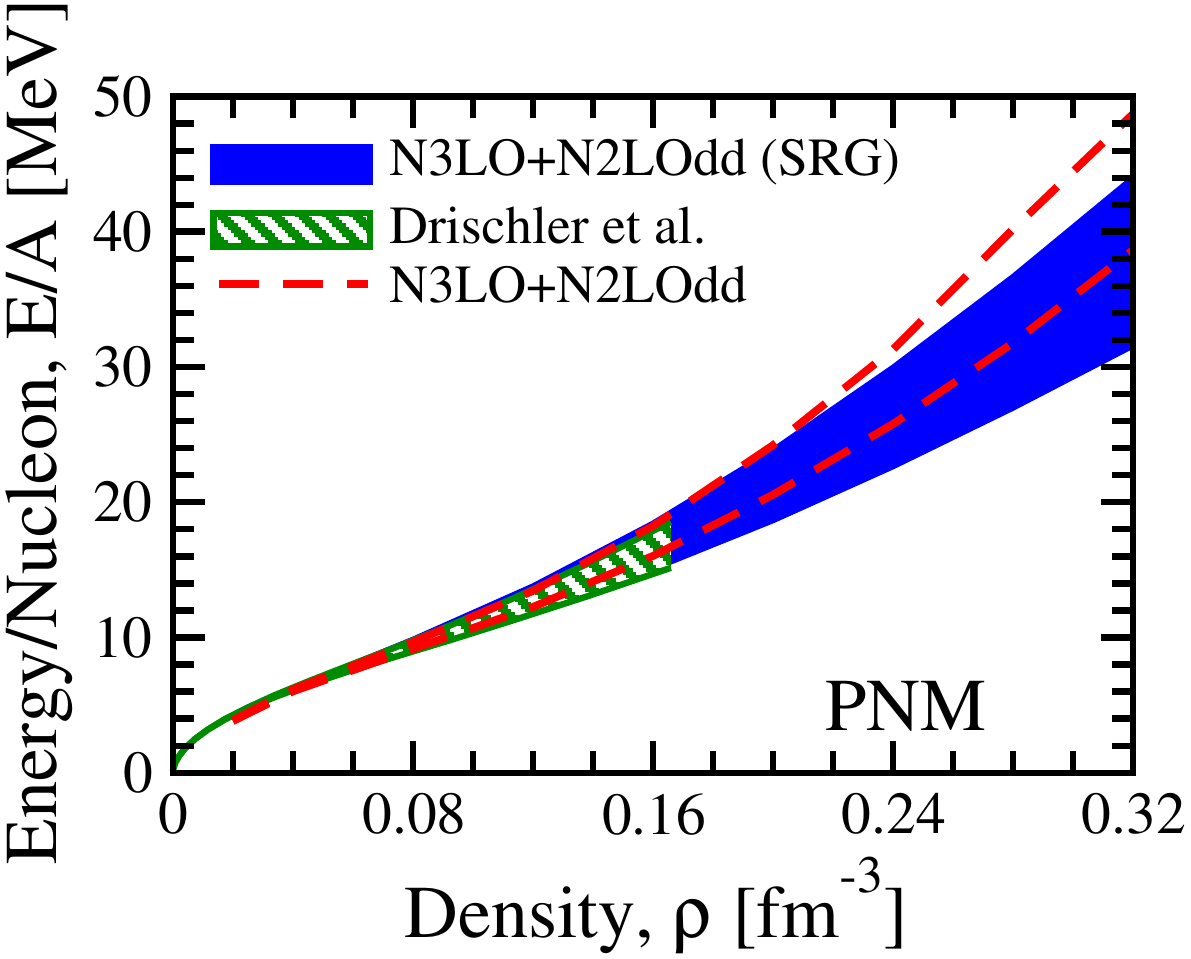}
\caption{(Color online) PNM energy per nucleon as a function of density at $T=0$ MeV including errors from LECs. Red-dashed lines show the calculation for the 2B N3LO plus the  density-dependent 2NF obtained with a correlated momentum distribution and a full regulator. The blue solid band corresponds to the calculation where a SRG evolution is applied on the 2B N3LO potential down to $\Lambda_\mathrm{SRG}=2.0$ fm$^{-1}$. The hatched band shows results obtained by Drischler \emph{et al.} \cite{Hebeler2010Jul,Drischler2014} in a perturbative calculation up to second-order with the same SRG-evolved input 2B potential and a density-dependent 2NF obtained from 3NFs at N2LO. The bands reflect the uncertainty on the underlying LECs, as explained in the text.}
\label{pnm_Kai_comp}
\end{center}
\end{figure}

Following the work presented by Hebeler \emph{et al.} in Refs.~\cite{Hebeler2010Jul,Hebeler2013Jul}, we show the LEC-induced error band in the PNM energy calculation in Fig.~\ref{pnm_Kai_comp}. 
We present three sets of calculations. First, we show the results obtained with the 2B N3LO  Entem-Machleidt force \cite{Entem2003} and the associated density-dependent 2NF with a full regulator and a correlated momentum distribution (dashed lines). In the second set of results, the 2B has been evolved via a similarity renormalization group (SRG) transformation down to a cutoff of $\Lambda_\textrm{SRG}=2.0$ fm$^{-1}$ \cite{Bogner2010}. The associated density-dependent 2B force is consistently calculated with a cutoff at $\Lambda_\textrm{3NF}=2.0$ fm$^{-1}$. We note that no induced 3NF generated in the SRG process are considered in this approach. Finally, the hatched band is obtained from a calculation with the same SRG-evolved 2NF plus a density-dependent force obtained from the 3NFs at N2LO. In this latter case the many-body results are obtained in a perturbative many-body calculation up to second order \cite{Hebeler2010Jul,Drischler2014}. The band in all three cases reflects the uncertainty associated to the LECs.

The width of the band is indicative of the systematic uncertainties in the calculations. We observe that in all cases the spread in results due to theoretical uncertainties in the LECs ranges from less than $1$ MeV at sub saturation densities to about $4$ MeV at $\rho_0$. As expected, the error band increases with density and at $2\rho_0$ the width becomes almost $10$ MeV wide. Error quantification is one of the major advantages of chiral EFT-based potentials. 

A complete error propagation scheme should also consider variations on regulators and cutoff constants \cite{Coraggio2013,Krueger2013}. Whereas a systematic study lies beyond the scope of our analysis, we note that the agreement between the results based on a bare 2NF and a SRG-evolved force indicates that these additional error sources are not significant up to about saturation density. Looking at Fig.~\ref{pnm_Kai_comp}, we indeed observe that all three bands overlap very well up to saturation density, confirming results already presented in Ref.~\cite{Hebeler2013Jul}. In this sense,  the theoretical uncertainties in the neutron matter equation of state are well under control. This agrees with indications from  perturbative calculations \cite{Hebeler2010Jul,Krueger2013,Coraggio2013}. If we now turn to double saturation density, the difference between the dashed line and the full band is about $5 \mbox{-}7$ MeV wide. The fact that this is smaller than the total width of the band pushes towards the idea that the uncertainty in the LECs dominates the error of the many-body calculations at high densities. However, other sources of error must be investigated, such as missing three- and four-body forces at higher order in the chiral expansion, as well as missing induced forces in the SRG-evolved calculation. Furthermore, the variation of $c_1$ and $c_3$ should be consistently considered in both the 2B and 3B sector \cite{Coraggio2013}.

\begin{figure}[t]
\begin{center}
\includegraphics[width=0.45\textwidth]{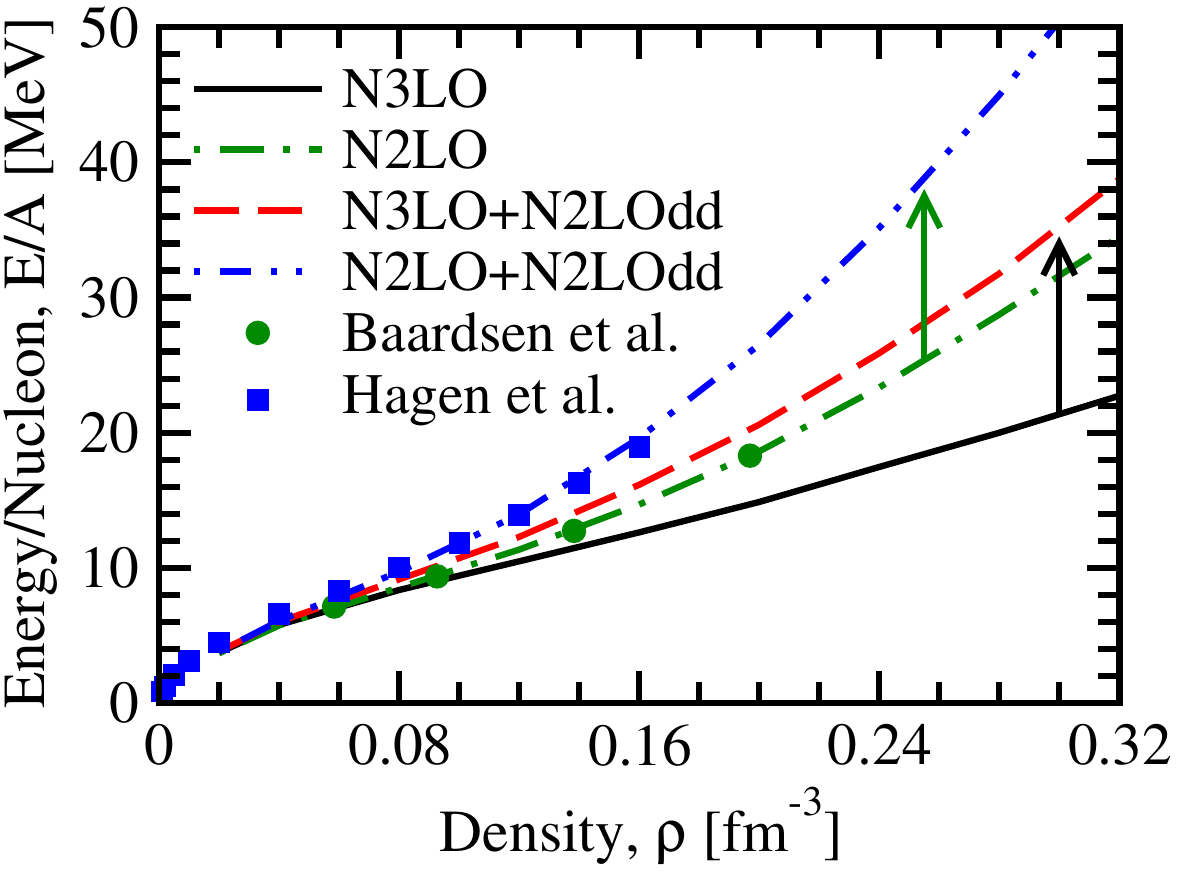}
\caption{
(Color online) PNM energy per nucleon as a function of density at $T=0$ MeV for different chiral interactions. The black-solid line and the green-dot-dashed lines show results with 2NF only, with either N3LO or N2LOopt potentials. The red-dashed line corresponds to calculation performed with 2B N3LO plus the N2LO density-dependent force obtained in the correlated version with a full regulator. The blue-double-dot-dashed line is obtained from a 2B N2LOopt potential and a density-dependent 2B force with the corresponding $c_1,c_3$ of N2LOopt \cite{Ekstroem2013}. The arrows connect the curves obtained with the N2LO (green-dot-dashed) and with the N3LO  (black-solid) 2NFs with the respective curves including 3NFs. The symbols are data from the pp-hh coupled cluster calculations of Baardsen \emph{et al.} \cite{Baardsen2013} and of Hagen \emph{et al.} \cite{Hagen2014}. }
\label{pnm_n3lo_n2lo}
\end{center}
\end{figure}

Finally, we present a series of results to discuss the consistency within the chiral expansion associated to the many-body results. We note that from this point on we do not provide any error analysis. To be consistent in the order of the chiral expansion, and following our previous study in SNM \cite{Carbone2013Oct}, we perform PNM calculations at N2LO using the newly optimized N2LOopt 2NF of Ref.~\cite{Ekstroem2013}. We construct a corresponding density-dependent 2NF associated to the N2LO 3B interaction with the corresponding $c_1$ and $c_3$ constants. This density-dependent force is obtained from a fully correlated and regulated average. The results obtained with 2NF only and with the density dependent 2B forces are presented in Fig.~\ref{pnm_n3lo_n2lo}.

We find that, whether or not we include density-dependent 2NFs, results obtained with N2LOopt are more repulsive than the N3LO calculations throughout the whole density regime. In the 2B only case, the repulsion is around $\sim2$ MeV at $\rho_0=0.16$ fm$^{-3}$ and grows with density, up to $\sim10$ MeV at $2\rho_0$. When the 3B force is included, the difference between the N2LO and N3LO results is a bit larger. At the highest density considered here, a repulsive effect of $\sim15$ MeV is found. We note, however, that the effect of including 3NFs is very similar in both cases. This is indicative of the similarity between the density-dependent 2B forces at N2LO, which are only modified by (a) a change in the LECs and (b) the differences in the correlated momentum distribution of each calculation. The underlying differences in the N2LO and N3LO 2B forces are more difficult to identify, and would ultimately be responsible for the large variation in the results  \cite{Carbone2013Oct}. We also note that, if the chiral expansion is valid, the error bands associated to LEC variations at N2LO should be larger than at N3LO \cite{Epelbaum2009}. 

In addition to the uncertainties associated to the NN force, the theoretical calculations we present are affected by the systematic uncertainty associated to the many-body method at choice. Figure \ref{pnm_Kai_comp}  indicates that neutron matter is well described perturbatively below saturation density. Furthermore, a comparison of our nonperturbative calculations with up-to-third-order results in the energy expansion presented by Coraggio \emph{et al.} in Ref.~\cite{Coraggio2013}, tests the PNM perturbative behavior up to densities above saturation density. One should therefore not expect large differences associated to the use of many-body techniques. We confirm that results are rather independent of the many-body method at choice by comparing our results to recent coupled-cluster calculations \cite{Baardsen2013,Hagen2014}. The latter account for particle-particle and hole-hole correlations in the equation of state from a coupled cluster perspective, rather than a SCGF one. We find that our calculations are in good agreement with the coupled-cluster results for the N2LOopt interactions. Calculations with harder interactions for both methods could provide a benchmark on their quality at the many-body level. Moreover, comparisons with other many-body calculations (Monte Carlo, for instance) would provide indications of the importance of ladder-like correlations in the equation of state. 

\section{Conclusions}

We have presented calculations for SNM and PNM exploiting the recently extended SCGF method. We include consistently chiral 2B and 3B forces in the ladder approximation. To take into account the effect of 3NFs without the need to explicitly calculate the 3B dressed GF, we have used 1B and 2B effective interactions. These include reducible effects of 3B physics in both 1B and 2B interactions. The effective interactions are averages of the original 3B force performed with the use of dressed propagators. Unlike previous calculations, the results presented here include consistently dressed internal propagators in the construction of a correlated density-dependent 2B and 1B forces. The use of consistent 1B Green's functions in the averaging procedure is the main innovation of this paper.

 In fact, to evaluate the effective 2B operator, we have implemented the correlated SP momentum distribution function obtained consistently at each step of iteration in the SCGF approach. The density-dependent force has been calculated from the contraction of the 3B terms appearing at N2LO in the chiral expansion. Particular attention has been paid in considering all possible terms arising in the averaging, following previous works from Refs.~\cite{Hebeler2010Jul,JWHolt2010}. We have tested the modifications of partial waves when including the density-dependent force on top of the 2B N3LO force. Furthermore, we have analyzed different ways of evaluating the 3B force average. These include the use of an uncorrelated momentum distribution, as well as variations on the regulator function in the momentum integrals of the 3B chiral force. We find small differences between different averaging procedures. This validates results obtained with uncorrelated averages in the literature. The small differences between approximations can be ascribed to the availability of momenta when integrating over the third particle. 

We have subsequently analyzed the effect of including the correlated density-dependent 2B force in microscopic and macroscopic properties of SNM. The momentum distribution at different densities is rather insensitive to the inclusion of 3NFs. In addition, it is independent from the average procedure used in the construction of the contracted force. A similar independence has been observed for the total energy of SNM.  We find a small enhancement of the absolute values of the total energy in going from a lower to a higher availability of momentum states in the average. 

For the first time, we have presented results for PNM in the framework of the extended SCGF method. We find a repulsive effect associated to the inclusion of the contracted 3NFs in the entire density range. We have explored the dependence of our results on the underlying uncertainties associated to the $c_1$ and $c_3$ LECs in the 2B and the 3B nuclear force. By applying a low-momentum evolution on the 2B part, we have tested the perturbative nature of PNM. We have observed that the LECs uncertainty dominates at high densities, however other sources of error in the calculation must be taken into account in order to drive a sound conclusion over this aspect. We have also performed calculations for PNM exploiting the N2LOopt 2BF. This has allowed for a consistent comparison with infinite-matter coupled-cluster calculations. The agreement between both methods is very good, which indicates that many-body errors are under control. 

The work presented here has focused on the construction of correlated chiral density-dependent forces consistent with the many-body framework used. The dressed average takes into account the correlations which characterize the system in the specific conditions under study and therefore goes beyond previous uncorrelated calculations. Further improvements include two-body averages using fully dressed two-body potentials and systematic studies of the uncertainties associated to the interaction. We consider these a step forward in the consistent inclusion of 3NFs in infinite nuclear-matter calculations.

\appendix
\section{Correlated density-dependent two-body force at N2LO}
\label{app}

In this appendix, we provide explicit expressions for the density-dependent interaction obtained by applying the averaging procedure of Eq.~(\ref{dd3bf}) to a chiral N2LO 3NF. We follow the notation of Ref.~\cite{JWHolt2010}, but we highlight explicitly the effect of correlations and regulators in the one-body momentum integrations. In the limit of a a zero-temperature uncorrelated momentum distribution and an external regulator (as defined in the text), we recover the expressions of Ref.~\cite{JWHolt2010}.

Whatever the structure of the 2B density-dependent force, it can always be expressed in a generic form \cite{Erkelenz1971}. Consider, for instance, the most general form for the matrix elements of a two-nucleon potential which is charge independent, Hermitian and invariant under translation, particle exchange, rotation, space reflection and time reversal:
\beqn
V({\bf k},{\bf q})&=&V^s_c+\bm{\tau}_1\cdot\bm{\tau}_2V^v_c
\label{on-shell_vnn}
\\\nn 	&&
+[V^s_\sigma+\bm{\tau}_1\cdot\bm{\tau}_2V^v_\sigma]\bm{\sigma}_1\cdot\bm{\sigma}_2
\\\nn &&
+[V^s_{\sigma q}+\bm{\tau}_1\cdot\bm{\tau}_2V^v_{\sigma q}]\bm{\sigma}_1\cdot{\bf q} \, \bm{\sigma}_2\cdot{\bf q}
\\\nn &&
+[V^s_{SL}+\bm{\tau}_1\cdot\bm{\tau}_2V^v_{SL}]i(\bm{\sigma}_1+\bm{\sigma}_2)\cdot({\bf q}\times{\bf k})
\\\nn &&
+[V^s_{\sigma L}+\bm{\tau}_1\cdot\bm{\tau}_2V^v_{\sigma L}]\bm{\sigma}_1\cdot({\bf q}\times{\bf k})\bm{\sigma}_2\cdot({\bf q}\times{\bf k})\,.
\nn
\enqn
The subscripts denote the following: $c$ for the central term, $\sigma$ for the spin-spin term, $\sigma q$ for the tensor term, $SL$ for the spin-orbit term, and $\sigma L$ for the quadratic spin-orbit term. All contributions are presented in an isoscalar $V^s$ and isovector $V^v$ form. This expression is useful in identifying the different contributions of the density-dependent interaction which arise from contractions of the 3NF terms written in Eqs.~(\ref{tpe})-(\ref{cont}). Furthermore, this form is helpful in finding the partial-wave decomposition of the matrix elements  \cite{Erkelenz1971,Kaiser1997,JWHolt2010}. We only consider matrix elements which are diagonal in relative momenta, i.e. $|{\bf k}|=|{\bf k'}|=k$. The generalization of the previous expression to off-diagonal elements is possible. For non-diagonal momentum matrix elements, however, Eq.~(\ref{on-shell_vnn}) includes a further operatorial structure \cite{Erkelenz1971}, which complicates the partial-wave decomposition. We therefore extrapolate off-diagonal momentum matrix elements from diagonal ones, with the prescription $k^2\rightarrow(k^2+k'^2)/2$ as proposed in Ref.~\cite{JWHolt2010}.

\subsection{Symmetric nuclear matter}

We now present expressions for averaged matrix elements in SNM. PNM results are discussed in the following subsection. Evaluating the trace of Eq.~(\ref{dd3bf}) for the TPE contribution of Eq.~(\ref{tpe}) depicted in Fig.~\subref*{TPE-3B}, one finds  three contracted terms for the in-medium 2B interactions. These are represented in Figs.~\subref*{TPE-1}-\subref*{TPE-3}. The first term, Fig.~\subref*{TPE-1}, is an isovector tensor contribution.  This is analogous to a 1$\pi$ exchange, with an in-medium pion propagator:
\beq
\tilde V_\mathrm{TPE-1}^\mathrm{3NF}=\frac{g_A\,\rho_f}{2 F_\pi^4}
\frac{(\bm{\sigma}_1\cdot{\bf q})(\bm{\sigma}_2\cdot{\bf q})}{[q^2 + M_\pi^2]^2}
\bm{\tau}_1\cdot\bm{\tau}_2[2 c_1M_\pi^2+ c_3\,q^2]\,.
\label{tpe_dd_1}
\enq
where $\rho_f$ is the integral of the correlated momentum distribution function weighed by the regulator function,
\beq
\frac{\rho_f}{\nu}=\int \frac{{\d}{\bf p}_3}{(2\pi)^3}n({\bf p}_3)f(k,k,p_3)\,,
\label{rho_f}
\enq
where $\nu$ is the spin-isospin degeneracy of the system ($\nu=4$ for SNM, $\nu=2$ for PNM). If the regulator does not depend on the internal momentum, $p_3$, the integral reduces to the total density of the system, $\rho$, 
\beq
\frac{\rho_f}{\nu} \to \frac{\rho}{\nu} \times f(k,k)\,.
\label{rho_f}
\enq
Consequently, whenever the regulator is external, the value of this integral is independent of whether the momentum distribution is correlated or not. This comment is relevant for two other contributions [see Eqs.~(\ref{ope_dd_1}) and (\ref{cont_dd}) below]. In other words, half of the density-dependent contributions are insensitive to correlations if external regulators are used.

The second term, Fig.~\subref*{TPE-2}, is also a tensor contribution to the in-medium NN interaction. It adds up to the previous term and contributes to $V^v_{\sigma q}$. This term includes vertex corrections to the 1$\pi$ exchange due to the presence of the nuclear medium as follows:
\beqn
\nn &&
 \tilde V_\mathrm{TPE-2}^\mathrm{3NF}= \frac{g_A^2}{8\pi^2F_\pi^4}
\frac{\bm{\sigma}_1\cdot{\bf q}\bm{\sigma}_2\cdot{\bf q}}{q^2 + M_\pi^2} \bm{\tau}_1\cdot\bm{\tau}_2
\\\nn  && \,\,\,
\times\Big\{-4c_1M_\pi^2\left[\Gamma_1(k)+\Gamma_0(k)\right]
\\\nn && \qquad
-(c_3+c_4)\left[q^2(\Gamma_0(k)+2\Gamma_1(k)+\Gamma_3(k))+4\Gamma_2(k)\right]
\\ && \qquad
+4c_4{\cal I}(k)\Big\}\,.
\label{tpe_dd_2}
\enqn
We introduce the functions $\Gamma_0(k)$, $\Gamma_1(k)$, $\Gamma_2(k)$, $\Gamma_3(k)$ and ${\cal I}(k)$, which are integrals over a single pion propagator:
\beq
\Gamma_0(k)=\int\frac{{\d}{\bf p}_3}{2\pi}n({\bf p}_3)
\frac{1}{[{\bf k}+{\bf p}_3]^2 + M_\pi^2}f(k,k,p_3)\,,\,\,\,\,\
\label{gamma0}
\enq
\beq
\Gamma_1(k)=\frac{1}{k^2}\int\frac{\d{\bf p}_3}{2\pi}n({\bf p}_3)
\frac{{\bf k}\cdot{\bf p_3}}{[{\bf k}+{\bf p}_3]^2 + M_\pi^2}f(k,k,p_3)\,,
\label{gamma1}
\enq
\beq
\Gamma_2(k)=\frac{1}{2k^2}\int\frac{\d{\bf p}_3}{2\pi}n({\bf p}_3)
\frac{p_3^2k^2-({\bf k}\cdot{\bf p_3})^2}{[{\bf k}+{\bf p}_3]^2 + M_\pi^2}f(k,k,p_3)\,,
\label{gamma2}
\enq
\beq
\Gamma_3(k)=\frac{1}{2k^4}\int\frac{\d{\bf p}_3}{2\pi}n({\bf p}_3)
\frac{3({\bf k}\cdot{\bf p_3})^2-p_3^2k^2}{[{\bf k}+{\bf p}_3]^2 + M_\pi^2}f(k,k,p_3)\,,
\label{gamma3}
\enq
\beq
{\cal I}(k)=\int \frac{{\d}{\bf p}_3}{2\pi}n({\bf p}_3)
\frac{[{\bf p}_3+{\bf k}]^2}{[{\bf p}_3+{\bf k}]^2 + M_\pi^2}f(k,k,p_3)\,.
\label{i_integral}
\enq
These integrals are formally equal to those presented in Ref.~\cite{JWHolt2010}, but differ in the use of a correlated momentum distribution and the explicit weighing of an internal regulator.

The last TPE contracted term, depicted in Fig.~\subref*{TPE-3}, includes in-medium effects for a 2$\pi$ exchange 2B term. This expression contributes to all operatorial structures of Eq.~(\ref{on-shell_vnn}). Specifically, it contributes to the scalar central term $V^s_c$,  to the isovector spin-spin $V^v_\sigma$ and tensor term $V^v_{\sigma q}$, to the spin-orbit in both isoscalar $V^s_{SL}$ and isovector form $V^v_{SL}$, and to the isovector quadratic spin-orbit term $V^v_{\sigma L}$ as follows:
\begin{widetext}
\beqn
\nn 
\tilde V_\mathrm{TPE-3}^\mathrm{3NF}=\frac{g_A^2}{16\pi^2F_\pi^4}\Big\{
&-& 12c_1M_\pi^2\big[2\Gamma_0(k)-G_0(k,q)(2M_\pi^2+q^2)\big]
\\\nn 
&-&c_3\big[8k_F^3-12(2M_\pi^2+q^2)\Gamma_0(k)
-6q^2\Gamma_1(k)+3(2M_\pi^2+q^2)^2G_0(k,q)\big] 
\\\nn
&+&4c_4 \bm{\tau}_1\cdot\bm{\tau}_2(\bm{\sigma}_1\cdot\bm{\sigma}_2\, q^2-\bm{\sigma}_1\cdot{\bf q}\bm{\sigma}_2\cdot{\bf q})
G_2(k,q)
\\\nn
&-&(3c_3+c_4\bm{\tau}_1\cdot\bm{\tau}_2)\,i(\bm{\sigma}_1+\bm{\sigma}_2)\cdot({\bf q}\times{\bf k})
\big[2\Gamma_0(k)+2\Gamma_1(k)-(2M_\pi^2+q^2)(G_0(k,q)+2G_1(k,q))\big]
\\\nn 
&-&12c_1M_\pi^2\,  i(\bm{\sigma}_1+\bm{\sigma}_2)\cdot({\bf q}\times{\bf k})
\big[G_0(k,q)+2G_1(k,q)\big]
\\
&+&4c_4\bm{\tau}_1\cdot\bm{\tau}_2\bm{\sigma}_1\cdot({\bf q}\times{\bf k})\bm{\sigma}_2\cdot({\bf q}\times{\bf k})
\big[G_0(k,q)+4G_1(k,q)+4G_3(k,q)\big]\Big\}\,.
\label{tpe_dd_3}
\enqn
Here, we have introduced the function $G_0(k,q)$, which is an integral over the product of two different pion propagators:
\beqn
G_{0,\star,\star\star}(k,q)=
\int \frac{{\d}{\bf p}_3}{2\pi}n({\bf p}_3)
\frac{\{p_3^0,p_3^2,p_3^4\}}{\big[[{\bf k}+{\bf q}+{\bf p}_3]^2+M_\pi^2\big]\big[[{\bf p}_3+{\bf k}]^2+M_\pi^2\big]}f(k,k,p_3)\,.\qquad\,\,\,\,\,
\label{G_0} 
\enqn
The starred $G$ functions, $G_{\star}(k,q)$, $G_{\star\star}(k,q)$ and $G_{1\star}(k,q)$ are auxiliary functions in that they are only used to define $G_1(k,q)$, $G_2(k,q)$, and $G_3(k,q)$ as follows:
\beqn
G_1(k,q)&=&\frac{\Gamma_0(k)-(M_\pi^2+k^2)G_0(k,q)-G_\star(k,q)}{4k^2-q^2}\,,
\label{G_1} \\
G_{1\star}(k,q)&=&\frac{3\Gamma_2(k)+k^2\Gamma_3(k)-(M_\pi^2+k^2)G_\star(k,q)-G_{\star\star}(k,q)}{4k^2-q^2}\,, 
\label{G_1star}\\
G_2(k,q)&=&(M_\pi^2+k^2)G_1(k,q)+G_\star(k,q)+G_{1\star}(k,q)\,,
\label{G_2} \\
G_3(k,q)&=&\frac{\Gamma_1(k)/2-2(M_\pi^2+k^2)G_1(k,q)-2G_{1\star}(k,q)-G_\star(k,q)}{4k^2-q^2}\,.
\label{G_3}
\enqn
\end{widetext}
At this point, we note that all $\Gamma_x$ and $G_x$ functions need to be evaluated numerically within the correlated approach. Special care is needed in the treatment of the high-momentum components.

Integrating Eq.~(\ref{dd3bf}) for the OPE 3NF term, given in Eq.~(\ref{ope}), yields two contributions. The first one, Fig.~\subref*{OPE-1}, is a tensor contribution which reads as a vertex correction to a 1$\pi$ exchange NN term,
\beq
\tilde V_\mathrm{OPE-1}^\mathrm{3NF}=-\frac{c_D\,g_A\,\rho_f}{8\,F_\pi^4\,\Lambda_\chi}
\frac{(\bm{\sigma}_1\cdot{\bf q})(\bm{\sigma}_2\cdot{\bf q})}{q^2 + M_\pi^2}
(\bm{\tau}_1\cdot\bm{\tau}_2)\,.
\label{ope_dd_1}
\enq
As for the $\tilde V_\mathrm{TPE-1}^\mathrm{3NF}$ term, $\tilde V_\mathrm{OPE-1}^\mathrm{3NF}$ is proportional to $\rho_f$ and contributes only to the isovector tensor term, $V^v_{\sigma q}$.

The second term arising from the 3NF OPE is depicted in Fig.~\subref*{OPE-2}. It defines a vertex correction to the short-range contact NN interaction.  This in-medium interaction is formed of a variety of contributions: a central scalar $V^s_c$, a  spin-spin $V^v_\sigma$, a tensor $V^v_{\sigma q}$ and a quadratic spin-orbit $V^v_{\sigma L}$ term. It reads as follows:
\beqn
\nn 
\tilde V_\mathrm{OPE-2}^\mathrm{3NF}=\frac{c_Dg_A}{16\pi^2F_\pi^4\Lambda_\chi} &\Big\{&
\Big[ \left[ \Gamma_0(k)+2\Gamma_1(k)+\Gamma_3(k) \right]
\\\nn 
&\times&\left[  \bm{\sigma}_1\cdot\bm{\sigma}_2\Big(2k^2-\frac{q^2}{2}\Big) \right.
\\\nn 
&& +(\bm{\sigma}_1\cdot{\bf q}\,\bm{\sigma}_2\cdot{\bf q})\Big(1-\frac{2k^2}{q^2}\Big) 
\\\nn && \left.
-\frac{2}{q^2}\bm{\sigma}_1\cdot({\bf q}\times{\bf k})
\bm{\sigma}_2\cdot({\bf q}\times{\bf k})\right]
\\ 
&&+2\Gamma_2(k)(\bm{\sigma}_1\cdot\bm{\sigma}_2)\Big](\bm{\tau}_1\cdot\bm{\tau}_2) \nn \\
&+&6{\cal I}(k)\Big\}\,.
\label{ope_dd_2}
\enqn

The last density-dependent contribution, shown in Fig.~\subref*{contact}, arises from a contraction of the contact 3NF term given in Eq.~(\ref{cont}). This yields a scalar central contribution to the in-medium NN interaction, proportional to $\rho_f$,
\beq
\tilde V_\mathrm{cont}^\mathrm{3NF}=-\frac{3 c_E\rho_f}{2 F_\pi^4\Lambda_\chi}\,.
\label{cont_dd}
\enq
 We note that, since this is a momentum independent term, it will contribute only to $S$ partial waves. 
 
We stress once again that the in-medium 2B interaction terms, Eq.~(\ref{tpe_dd_1}), Eq.~(\ref{tpe_dd_2}), Eq.~(\ref{tpe_dd_3}), Eq.~(\ref{ope_dd_1}), Eq.~(\ref{ope_dd_2}) and Eq.~(\ref{cont_dd}), are formally the same as those obtained by the authors in Ref.~\cite{JWHolt2010}. The difference lies in the intermediate integrals, which are performed using the correlated momentum distribution function $n({\bf p}_3)$, and a full regulator function $f(k,k',p_3)$. Because of their self-consistent nature, these density-dependent matrix elements are computed numerically. 

\subsection{Pure neutron matter}

In the case of PNM, the evaluation of Eq.~(\ref{dd3bf}) is simplified as the trace over isospin is trivial. Consequently the exchange operator of Eq.~(\ref{perm_op}) reduces only to the momentum and spin part. It can then be proved that the terms proportional to $c_4, c_D, c_E$ go to zero in PNM when using a non-local regulator \cite{Tolos2008,Hebeler2010Jul}. 

In summary, the only density-dependent contributions which are non-zero in PNM are those proportional to $c_1$ and $c_3$ in Eq.~(\ref{tpe}). Formally, the density-dependent  terms obtained in PNM will only differ with respect to those in SNM by some prefactors. This is because of the disappearance of the isospin-exchange operation in PNM matrix elements. To account for the fact that Eq.~(\ref{rho_f}) has a different degeneracy in PNM, we need to replace $\rho_f \rightarrow 2\rho_f$ in the $\tilde V_\mathrm{TPE-1}^\mathrm{3NF}$ contribution of Eq.~(\ref{tpe_dd_1}).  In addition, the isovector tensor terms $\tilde V_\mathrm{TPE-1}^\mathrm{3NF}$ and  $\tilde V_\mathrm{TPE-2}^\mathrm{3NF}$, given in Eqs.~(\ref{tpe_dd_1})-(\ref{tpe_dd_2}), change prefactors according to:
\beqn
\tilde V_\mathrm{TPE-1}^\mathrm{3NF} &&: \bm{\tau}_1\cdot\bm{\tau}_2 \rightarrow \frac{1}{2}\bm{\tau}_1\cdot\bm{\tau}_2\,, 
\label{pnm_tpe_1} \\
\tilde V_\mathrm{TPE-2}^\mathrm{3NF}&&: \bm{\tau}_1\cdot\bm{\tau}_2 \rightarrow 
\frac{1}{4}(\bm{\tau}_1\cdot\bm{\tau}_2-2)\,.
\label{pnm_tpe_2}
\enqn
Further, the isoscalar part of the density-dependent potential appearing in $\tilde V_\mathrm{TPE-3}^\mathrm{3NF}$ becomes 
\beq
\tilde V_\mathrm{TPE-3}^\mathrm{3NF}: 1 \rightarrow \frac{1}{3}\,.
\label{pnm_tpe_3}
\enq
With these changes, the density-dependent 2B matrix elements can be used in PNM calculations. We note that, as in the SNM case, the fully correlated matrix elements are very similar to the uncorrelated ones presented previously in the literature \cite{Hebeler2010Jul,JWHolt2010}.

\begin{acknowledgments}
We are grateful to C. Barbieri, K. Hebeler, A. Lovato,  A. Cipollone, V. Som\`a and J. W. Holt for very fruitful and enlightening discussions. We thank C. Drischler and M. Hjorth-Jensen for providing us the data of Fig.~\ref{pnm_Kai_comp} and Fig.~\ref{pnm_n3lo_n2lo} respectively.  This work is supported by the Consolider Ingenio 2010 Programme CPAN CSD2007-00042, Grant No. FIS2011-24154 from MICINN (Spain) and Grant No.~2014SGR401 from Generalitat de Catalunya  (Spain); and by STFC, through Grants ST/I005528/1, ST/J000051/1 and ST/L005816/1. A. Carbone acknowledges the support by the SUR of the ECO from Generalitat de Catalunya and the DFG through Grant SFB 634. Partial support comes from ``NewCompStar" COST Action MP1304.
\end{acknowledgments}

\bibliographystyle{apsrev4-1}
\bibliography{biblio}

\end{document}